\input epsf
\documentstyle{elsevier}
\font\tencop=bbm12
\font\sevencop=bbm9 \font\fivecop=bbm7
\newfam\copfam\scriptscriptfont
\copfam=\fivecop\textfont\copfam=\tencop\scriptfont\copfam=\sevencop
\def\cop{\fam\copfam\tencop}
\def\IM{\mathop{\Im m}\nolimits}
\def\RE{\mathop{\Re e}\nolimits}
\def\Z{{\cop Z}}
\def\R{{\cop R}}
\def\Q{{\cop Q}}
\def\zero{\vert 0\rangle }
\def\sn{\mathop{\rm sn}\nolimits }
\def\dn{\mathop{\rm dn}\nolimits }
\def\cn{\mathop{\rm cn}\nolimits }
\begin{document}
\begin{frontmatter}
\title{The dyon spectra of finite gauge theories}
\author{Frank Ferrari}
\address{Laboratoire de Physique Th\'eorique de l'\'Ecole Normale
Sup\'erieure\thanksref{UPR}}
\address{24 rue Lhomond, 75231 Paris Cedex 05, France\\
{\tt Frank.Ferrari@lpt.ens.fr}}
\thanks[UPR]{Unit\'e Propre de Recherche 701 of CNRS associated with
the \'Ecole Normale Sup\'erieure and with the University of Paris XI.}
\begin{abstract}
It is shown that all the $(p,q)$ dyon bound states exist and are unique in
$N=4$ and $N=2$ with four massless flavours 
supersymmetric SU(2) Yang-Mills theories, where
$p$ and $q$ are any relatively prime integers. The proof
can be understood in the context of field theory alone,
and does {\em not} rely on any duality assumption.
We also give a general physical argument showing that these theories
should have at least
an exact $\Gamma (2)$ duality symmetry, and then deduce in particular the
existence of the $(2p,2q)$ vector multiplets in the $N_f=4$ theory.
The corresponding massive theories are studied in parallel, and
it is shown that though in these cases the spectrum is no longer self-dual
at a given point on the moduli space, it is still in perfect
agreement with an exact S duality.
We also discuss the interplay between our results and both the
semiclassical quantization and
the heterotic-type II string-string duality conjecture.
\endkeyword
\end{abstract}
\end{frontmatter}
\setbox1=\hbox{LPTENS-96/67}
\setbox2=\hbox{\tt hep-th@xxx/9702166}
\setbox3=\vtop{\null\epsfbox{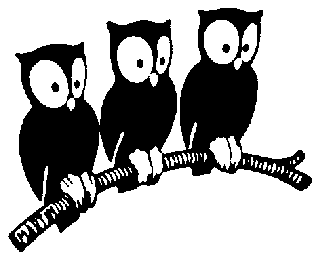}}
\makeatletter
\global\@specialpagefalse
\def\@oddhead{\hskip -1.2cm\box3\hfill\vbox{\vskip -1cm\box1\box2\vskip 1.4cm}}
\let\@evenhead\@oddhead
\section{Introduction}
In the late 70's, Montonen and Olive discovered that
a given four dimensional gauge theory might be described using different
sets of elementary fields \cite{MO}. The fields in one of this set
create the standard perturbative spectrum of the theory 
(photon, W bosons, quarks, \dots ), while the fields in the
``dual'' set correspond to the solitonic states (dyons).
An example of such a phenomenon was observed even before in some
two dimensional theories \cite{COL}.

Since then, a natural and exciting question 
arose: can we find theories where
the description in terms of the solitonic states is the same (same
field content and symmetries in the lagrangian)
as the description in terms of the perturbative states?
Answering this question requires to understand the strong coupling
behaviour of the gauge theories under study, since exchanging
electrically charged perturbative states with magnetically charged
solitonic states amounts to inverting the gauge coupling constant
due to the Dirac quantization condition. This was out of reach
in the 70's, though some important progress was made. First, it
was realized that the natural arena for electric-magnetic duality
was Yang-Mills theories with extended supersymmetries \cite{WO}.
Second, it was pointed out that in order to have dyons of spin 1,
which could be the dual of the W bosons, $N=4$ supersymmetry seemed
to be required \cite{OS}. Actually, an exact electric-magnetic duality, 
combined with the Dirac quantization condition,
implies that the $\beta $ function must vanish. This is indeed the only case
where the electric coupling $g$ and the magnetic coupling $1/g$,
being constant,
have the same behaviour under the renormalization group flow.
The $N=4$ theory is known to be conformally invariant
at the quantum level, perturbatively \cite{BO} as well
as non-perturbatively \cite{BOE}. At the perturbative level, this property 
is shared with some $N=2$ theories \cite{BO2}, which are believed to be
finite even non-perturbatively at least when the rank of the gauge group is one.
The latter theories are strongly believed to have, 
together with the $N=4$ theory,
an exact electric-magnetic duality symmetry; we will study them below.

There are various ways to test these duality conjectures. One of them,
advocated in \cite{SWII}, consists of looking at the (hyper)elliptic curve
from which the low energy physics can be deduced, and check whether
some sensible duality transformations can be defined in order to
insure exact electric-magnetic duality symmetry. Another approach
has been to look at the free energy \cite{FE}. But the most popular
test certainly is to determine the content of a particular 
sector of the Hilbert space,
the BPS sector, and to check whether it is compatible 
with duality \cite{SZ,Segal}.
To study solely the BPS states is not too restrictive. Actually,
all the perturbative and known stable solitonic states are BPS states.
Moreover, this is in general a very accurate test. For instance,
it is well known that the BPS spectra of the asymptotically free
$N=2$ theories certainly do not have the same symmetries as the
associated (hyper)elliptic curve.
There are also some technical reasons to study the BPS spectra.
An exact quantum mass formula
is known for these states, stemming from the fact that they lie in
small representations of the supersymmetry algebra and thus saturate
the Bogomoln'yi bound \cite{BOG}.
The semiclassical quantization  
can give reliable results when one can go continuously from weak coupling
to strong coupling without altering the stability of the states,
which is the case for instance in the $N=4$ theory.
And, what is maybe their most intriguing property, it seems that
the BPS spectra can be computed completely once one knows the low
energy structure of the theory, that is all the information about
these spectra seems to be contained, in a very hidden way, in the
(hyper)elliptic curve.

In this paper, I will study in detail the BPS spectra of two 
$N=2$ theories having zero $\beta $ function. The first one will be
the SO(3) gauge theory with one flavour of bare mass $m$, 
which reduces to the $N=4$ theory when $m=0$,
and the second one will be the SU(2) gauge theory with four hypermultiplets
whose bare masses will be taken to be $m_1=m_2=m/2$, $m_3=m_4=0$.
The fact that these seemingly very different theories can be treated
in parallel stems from the fact that their low energy effective
action are formally identical. I will use a method whose spirit
originated in \cite{FB,BF} and which is completely
understandable in the framework
of non perturbative field theory \`a la Seiberg-Witten \cite{SWI,SWII}.
This will lead to a natural and rigorous proof that
the spectra of both the $N=4$ and the $N=2$ theories are self-dual when the
bare masses are zero. We will also see how the duality
can still work in the massive theories though, as it will be shown,
the spectra are no longer in general self-dual {\em at a given point} in the
moduli space. 

The plan of the paper is as follows. In Section 2 some generalities
on the theories under study are recalled. Particular emphasis is put
on the quantum numbers carried by the solitonic states, as
suggested by a semiclassical analysis, and it is explained why they
are compatible with an exact duality symmetry.
In Section 3, after 
a short presentation of the Seiberg-Witten curves whose exactitude
will be our unique, very mild hypothesis, some of the physical ideas
which are at the basis of the present work are presented. This leads to a
limpid understanding of {\em why} states having any magnetic charges
must exist in the theory, and also gives a flavour of what the spectra in the
massive cases look like.
It appears that in some regime, the magnetic charge is quantized much
as if it were a periodic variable, while only one value of the
electric charge is allowed.
Though the spectrum does not appear
in general as being self-dual at a given point in moduli space, it is
argued that it is nevertheless perfectly compatible with an exact
S duality of the massive theories.
It is also pointed out that semiclassical 
reasonings might be able to account for the curious disappearance
of all but very few states in some regions of moduli space, a
phenomenon first discovered in \cite{FB} at strong coupling.
In order
to prove the complete SL$(2,\Z )$ invariance of the spectra, the 
consideration of theories with non zero bare $\theta $ angle is needed.
This is done in Section 4. We also discuss the appearance of
superconformal points in some particular theories.
In Section 5, the uniqueness of the states is proven, as required
by duality. In Section 6, a general presentation of the curves of
marginal stability, across which the BPS spectra may be discontinuous,
is given. These curves are then used in Section 7 to rigorously
establish the existence of all the $(p,q)$ states, for $p$ and $q$
relatively prime integers, in the massless theories. 
Finally, in Section 8, a general physical 
argument is presented which shows that 
the theories under study must have at least an exact 
$\Gamma (2)$ duality symmetry. In particular, the BPS spectra
of the {\em massless} theories must be invariant under the 
monodromy group $\Gamma (2)$ of the {\em massive} theories. We then deduce
that the vector $(2p,2q)$ states exist
in the massless $N_f=4$ theory. In Appendix A, the computation
of the Seiberg-Witten periods is presented, and in Appendix B
the curves of marginal stability are used to prove the existence of some
particular states required by duality in the massive theories.
\section{The quantum numbers of the BPS states}
As already noted in the Introduction, one easy way to check whether
a given theory may be, or cannot be, self-dual, is to look at the
quantum numbers carried by the states which are supposed to be transformed
into each other by an electric-magnetic rotation. For example, 
the W bosons must be the dual of a charged particle having
spin one in the one monopole sector. The very existence of such a spin one 
solitonic state is already a non trivial test of duality and was proven in
\cite{OS}. The main goal of this paper will be  to extend this kind of result
to all the electric and magnetic quantum numbers. In addition to the
electric and magnetic charges, there is another abelian quantum
number, which we will call the $S$ charge, which plays an important
r\^ ole, since it also appears in the central charge of the supersymmetry
algebra and thus in the BPS mass formula. The aim of this Section
is to explain the relations that may exist between these quantum
numbers, and also to discuss the way one should compute the
$S$ charges.
\subsection{Field content and BPS mass formula}
\subsubsection{The SO(3) theory}
The theory whose matter content is one $N=2$ adjoint ``quark'' hypermultiplet
will be called hereafter the SO(3) theory. This matter multiplet
consists in two chiral $N=1$ superfields $(Q,\tilde Q)$, which
correspond in terms of ordinary fields to
one Dirac spinor $\psi $ and two complex scalars $q$ and 
$\tilde q$ in the adjoint representation of the gauge group. 
The microscopic lagrangian also contains one Yang-Mills $N=2$ multiplet
(one vector field $A_\mu $, two Majorana spinors
$\lambda _1$ and $\lambda _2$ and one complex scalar $\phi $).
When the bare mass $m$ of the matter multiplet is zero, we recover
the $N=4$ theory. When $m \not = 0$, the moduli space of vacua is
a Coulomb branch where the gauge symmetry is spontaneously broken
down to U(1) by a non zero Higgs vacuum expectation value, 
$\langle\phi\rangle = {1\over 2} a \sigma _3$. When $m=0$ and
$N=4$, we have other equivalent Coulomb branches, permutted by the
SU(4)$_R$ symmetry, corresponding to the other scalar fields having
a vev. In addition to the U(1)s associated to the electric $Q_E$
and magnetic $Q_M$ charges, the theory has an abelian U(1) global
symmetry $Q\rightarrow e^{i\alpha}\, Q$, $\tilde Q\rightarrow
e^{-i\alpha}\,\tilde Q$, with corresponding charge $S$. 
Classically, the central charge of the
supersymmetry algebra is
\begin{equation}
\label{cccl}
Z_{\mathrm cl} = {a\over g}\,\Bigl( i Q_M - Q_E \Bigr) + {m\over\sqrt{2}}\, S,
\end{equation}
where $g$ is the gauge coupling constant.
Semiclassically, the electric charge $Q_E$ is $n_e g$, $n_e$ integer,
up to terms coming from CP violation (like a standard Witten term
proportional to the bare $\theta $ angle). The magnetic charge
is given by the Dirac quantization condition, $Q_M = 4\pi n_m /g$
for an integer $n_m$. We will loosely call in the following the
integers $n_e$ and $n_m$ the electric and magnetic charges.
The mass of a BPS state $(n_e,n_m,S)$ will then be
\begin{equation}
\label{BPSmass}
m=\sqrt{2}\,\vert Z\vert.
\end{equation}
For instance, an elementary quark $(1,0,1)$ has a tree level mass
$m=\sqrt{2}\,\vert a-m/\sqrt{2}\vert $.
The exact quantum formula for $Z$ can be found. It was shown in \cite{SWII} 
that it can be cast in the form
\begin{equation}
\label{cc}
Z= a_D n_m - a n_e + {m\over\sqrt{2}}\, s.
\end{equation}
Here $a_D$ is the dual variable of $a$ and can be, at least in principle,
computed very explicitly as a function of 
a gauge invariant coordinate $z$ on the moduli space ($z$ is
$\langle {\mathrm tr}\, \phi ^2\rangle$ up to a constant) \cite{SWII}.
At the tree level, we have $a_D=\tau a$, where $\tau =\theta /2\pi 
+ 4i \pi /g^2$ is a very convenient combination of the gauge coupling
$g$ and the theta angle $\theta $ which has a simple transformation
law under S duality. Note that $\tau $ is not renormalized since the
$\beta $ function is zero. Together with the mass $m$ and the
coordinate on the moduli space $z$, it is the only parameter
in the theory.

The constants $s$ in (\ref{cc})
are only indirectly related to the physical $S$
charges \cite{FER}, but can be computed very explicitly, see below.
\subsubsection{The $N_f=4$ theory}
This theory differs from the preceding one by its matter content.
Now we have four $N=2$ hypermultiplets $(Q_j,\tilde Q_j)$,
$j=1,\ldots , 4$, transforming in the fundamental
representation of the gauge group SU(2). 
We will limit ourselves, for reasons that will become clear in Section 3,
to the case where two of these hypermultiplets are massless, say
$m_3=m_4=0$, and the other two have identical masses $m_1=m_2=m/2$.
The global non abelian symmetry of the theory is then ${\mathrm Spin(4)}\times
{\mathrm SU(2)}={\mathrm SU(2)}\times {\mathrm SU(2)}\times {\mathrm SU(2)}$
(concerning the occurence of the universal cover Spin(4)
of SO(4) instead of SO(4) itself
for the symmetry associated with the two massless hypermultiplet, see
the next subsection). This shows that some states are likely to come
in doublets in this theory, as we will see later.

The moduli space of vacua has again a Coulomb branch where dyons can
exist, in addition to Higgs branches which we will not study.
The formula (\ref{cccl}) is still valid along the Coulomb branch, but
now we will have in general $Q_E=gn_e/2 + \cdots $, $n_e$ integer,
instead of $Q_E=gn_e+ \cdots $,
because the gauge group is really SU(2) and no longer SO(3) in this
theory. The elementary quarks have $n_e = \pm 1$ for instance.
Moreover, the $S$ charge will correspond to the transformations
$Q_j\rightarrow e^{i\alpha}\, Q_j$, $\tilde Q_j\rightarrow
e^{-i\alpha}\,\tilde Q_j$ for the quarks $j=1$ and $j=2$ simultaneously.

At a given point on the moduli space of vacua, the parameters 
of the theory will be $m$ and a generalized unrenormalized
gauge coupling $\tilde\tau =\theta /\pi + 8i\pi /g^2$.
Note the difference between this definition and the corresponding one 
in the SO(3) theory. The advantage of this choices will become apparent
in Section 3.
Moreover, in order to keep the formula (\ref{cc}) in the same
form in the two theories, we will define here the parameter $a$ by
$\langle \phi\rangle = a\sigma _3$. These different sets of conventions
were already used in \cite{SWII} and allow to make the formal similarity
between the two theories under study very explicit, see Section 3.
\subsection{Results from semiclassical quantization}
From the semiclassical point of view, the solitonic states appear
as bound states of a supersymmetric quantum mechanics describing the
low energy dynamics of interacting dyons \cite{GAU,SEN,SZ}. 
The dynamical variables
of this supersymmetric quantum mechanics correspond to zero modes
of the elementary fields of the quantum field theory. 
In a $n_m$ monopole configuration, there are $4n_m$ real bosonic zero modes.
The four zero modes that are present for all $n_m \not =0$
correspond to the center of mass motion (three modes) and to the
global electric charge (one periodic mode).
The other $4n_m-4$ zero modes describe the relative motions and
electric charges. To these bosonic zero modes are associated some
fermionic one. From Callias' index theorem \cite{CAL} we know that the two 
adjoint Majorana spinors of the $N=2$ vector multiplet will give $2n_m$
complex zero modes $\lambda ^a_{\pm }$, $1\leq a\leq 2n_m$,
carrying spin $S_z=\pm 1/2$.  
The other fermionic zero modes come from the matter fermions and differ
in the SO(3) and $N_f=4$ theory.
\subsubsection{The SO(3) theory}
There, the adjoint fermions in the matter multiplet will yield
$2n_m$ additional complex fermionic zero modes $\psi ^a_{\pm}$. These zero
modes carry spin $S_z=\pm 1/2$ 
and $S$ charge $+1$. The quantization of the four
$\lambda $ and $\psi $ fermionic zero modes corresponding to $a=1$ yields to
a spectrum of $2^4=16$ states. When $m=0$, this correspond to a full
short $N=4$ multiplet. This shows that monopoles $n_m=1$ may be the
dual of the W bosons. When $n_m\geq 2$, there are some additional 
zero modes. However it will become clear in the
following Sections that the $n_m\geq 2$ stable bound states that exist
in the theories under study all have the same quantum numbers as 
the $n_m=1$ monopoles (in some sense, they can be generated by a continuous 
deformation of the $n_m=1$ monopoles or of the elementary excitations). 
This means that for the stable states,
the additional zero modes play no r\^ole, and thus we will discard
them in the following.

When the mass $m$ is set to a non zero value, the $N=4$ multiplet
splits into three parts corresponding to states having the same 
$S$ charge and thus the same mass. More precisely, from the
sixteen states of the $N=4$ multiplet we get: eight states
corresponding to one $N=2$ vector multiplet, which may
be the dual to the $N=2$ W bosons $(n_e=1,n_m=0,S=0)$; 
four states corresponding to one-half
of a CPT self conjugate $N=2$ hypermultiplet (the other half is obtained
by quantization in the $-n_m$ monopole sector), which may be the dual
of one component of the adjoint elementary quark, say $(\pm 1,0,\pm 1)$;
and four states which may be the dual to $(\pm 1,0,\mp 1)$.
This is summarized in Table 1.
\begin{table}
\label{states1}
\caption{The sixteen states of a $N=4$ multiplet, with their spin
and $S$ charge. $S_0$ is the a priori
unknown $S$ charge of the state $\zero $. The $N=2$ content of the
$N=4$ multiplet is also indicated.}
\begin{tabular}{|p{3.5pc}|p{3.5pc}|p{4.5pc}|p{5.75pc}|
p{4.5pc}|p{3.5pc}|p{2.5pc}|}
\hline
\multicolumn{1}{|p{3.3pc}||}{N=2 content} 
& $S_z= -1$ & $S_z = -1/2$ & $S_z = 0$ & $S_z = 1/2$ & $S_z = 1$ &
\multicolumn{1}{||p{2.5pc}|}{$S$} \\
\hline
1/2 hyper & &$\lambda _-\zero $  &$\zero $, $\quad\quad $
$\lambda _+\lambda-\zero $  &
$\lambda _+\zero$  &  & $S_0$\\
\hline
vector    &$\psi _-\lambda _-\zero $  & $\psi _-\zero$, $\psi _-\lambda _+
\lambda _- \zero $  &$\psi _+\lambda _-\zero$, $\psi _-\lambda _+\zero $  &
$\psi _+\zero $, $\psi _+\lambda _+\lambda _-\zero $  &
$\psi _+\lambda _+\zero $  & $S_0 + 1$\\
\hline
1/2 hyper &  &$\psi _+\psi _-\lambda _-\zero $  &$\psi _+\psi _-\zero $,
$\psi _+\psi _-\lambda _+\lambda _-\zero $  &$\psi _+\psi _-\lambda _+\zero $
&  & $S_0 + 2$\\
\hline
\end{tabular}
\end{table}
\subsubsection{The $N_f=4$ theory}
There each of the four Dirac fermions belonging to the matter
multiplets $(Q_j,\tilde Q_j)$ 
will yield $n_m$ complex zero modes $\psi ^a_j$ \cite{CAL}
without spin ($\psi ^a_1$ and $\psi ^a_2$ carry one unit of $S$ charge,
while $\psi ^a_3$ and $\psi ^a_4$ have $S=0$). 
The only zero modes carrying spin are thus in this case
the $\lambda _{\pm}^a$. Quantization in the $n_m=1$ monopole sector
will thus lead to the spin content of a $N=2$ hypermultiplet, and such
states cannot be the dual of the W bosons. Moreover, we will see soon
that $n_m=1$ states transform non trivially under the action of the
flavour symmetry group, unlike the W bosons. Actually, the states
dual to the W will be $n_m=2$ states, which can both have the required
spin (because of the doubling of the number of $\lambda $-type
zero modes) and be singlets of the flavour group. 

Let us discuss first the case $m=0$. Then the (classical) flavour
symmetry group is SO(8), of which the $\psi _j$ corresponding
to the one monopole sector generate the Clifford algebra. This
yields a sixteen dimensional
(reducible) representation of Spin(8), the universal cover of SO(8).
As a $2\pi$ electric rotation acts 
non trivially on the $\psi _j$ (recall that the matter Dirac spinors 
are in the spin $1/2$ representation of the gauge group SU(2)),
these sixteen states are discriminated according to their electric
charge being even or odd.
We will obtain eight states
having an even electric charge and transforming in one spinor 
(irreducible) representation
of Spin(8) (${\bf 8_s}$), and eight states having an odd electric charge
transforming in the other spinor representation (${\bf 8_c}$). 
These multiplets are the candidates to be the duals of the
elementary quarks. They have the right spin content. However, the
elementary quarks transform in the vector representation $\bf 8_v$
of SO(8), which is not equivalent to $\bf 8_s$ or $\bf 8_c$.
This might appear prohibitive at first sight. However, the three
representations $\bf 8_v$, $\bf 8_c$ and $\bf 8_s$, though
inequivalent, are related by the outer automorphisms of SO(8).
This means that, by simply relabelling the elements of SO(8) in a way
that respects the group structure, one can permute the three representations.
Thus there is no obstruction at this level for the monopoles
$(2n,1)$, $(2n+1,1)$ and the quarks $(1,0)$ to be duals of each other,
but SL$(2,\Z )$ duality must be mixed with SO(8) triality \cite{SWII}.
The action of SL$(2,\Z )$ on the three representations
$({\bf 8_v},{\bf 8_c},{\bf 8_s})$ is easily obtained. For any
SL$(2,\Z )$ matrix $M$, consider its reduction modulo 2. You then
obtain an element of $\mathrm{SL}(2,\Z )/\sim $, where $\sim $
is the equivalence relation modulo 2. This group is isomorphic to the
group $S_3$ of permutations of three objects, and acts on the triplet
$({\bf 8_v},{\bf 8_c},{\bf 8_s})$. It is generated
by the transpositions $(213)$ which corresponds to the matrix ${}^t T$,
$(132)$ which corresponds to $T$ and $(321)$ which corresponds to
$S$, where as usual
\begin{equation}
T=\pmatrix{1 & 1 \cr 0 & 1 \cr },\quad {\mathrm and}\quad
S=\pmatrix{0 & 1 \cr -1 & 0 \cr }.
\end{equation}
What about the states with $n_m\geq 2$? There, on the one hand,
we have more $\psi $ zero modes
carrying flavour indices, and on the other hand we have more $\lambda $
zero modes carrying spin indices.
However, as in the SO(3) theory, we will see that the states having
$n_m\geq 2$ which are stable can be obtained continuously from the
$n_m=0,1$ states. Thus it will be enough for us to observe that we can indeed
construct $N=2$ hypermultiplets which are SO(8) spinors or vectors
for any $|n_m|\geq 1$ (the duals of the quarks), and $N=2$ vector multiplets
which are SO(8) singlets for any even $|n_m|\geq 2$ (the duals
of the W bosons).

Let us now go to the $m\not = 0$ theory. The mass term
breaks the flavour symmetry down from Spin(8) to
$\mathrm SU(2)_v\times Spin(4)= SU(2)_{v}\times SU(2)_{c}\times 
SU(2)_{s}$
(our notation will soon become clear).
The $\mathrm SU(2)_v$ factor comes
from the fact that the massive hypermultiplets have identical masses,
and we also have a Spin(4) symmetry from the two 
massless hypermultiplets. 
It is interesting to see how the Spin(8) multiplets of the massless
theory, both for the perturbative states and in the one monopole sector,
rearrange. On the one hand, the SO(8) vector multiplet of fundamental quarks 
splits into three flavour multiplets which are differentiated by their
$S$ charges and thus their physical masses. Noting a given BPS
multiplet by a triplet $(n_{e},n_{m},S)$, we have
two $\mathrm SU(2)_v$ doublets $(\pm 1,0,\pm 1)$ and $(\pm 1,0,\mp 1)$ and one
SO(4) vector $(\pm 1,0,0)$. On the other hand, the SO(8)
spinor multiplets of dyons states also splits into three
$\mathrm SU(2)_{v}\times SU(2)_{c}\times SU(2)_{s}$
multiplets as indicated in Table 2. It is not difficult to
determine how each of the three SU(2) factors acts on the representation
space of the SO(8) Clifford algebra. 
$(\psi _1\zero, \psi _2\zero)$ is a doublet of
$\mathrm SU(2)_v$ as can be seen by going back to the original field
variables. $(\psi_3\zero ,\psi_4\zero)$ and $(\zero ,\psi_3\psi_4\zero )$
are the basis for the two spinor representations of SO(4), and thus
form SU(2) doublets of two SU(2) factors which we call respectively
$\mathrm SU(2)_c$ and $\mathrm SU(2)_s$. This allows to determine the flavour
content in the monopole sector, as indicated in Table 2. 
One see that SL$(2,\Z )$ duality is now mixed with the permutation
of the three SU(2) factors $\mathrm (v,c,s)$, exactly in the same way 
SL$(2,\Z )$ duality was mixed with SO(8) triality in the massless case.
Actually one can readily check that the three SU(2), viewed as
subgroups of Spin(8), are permuted by the outer automorphisms
which also exchanged the $\bf 8_v$, $\bf 8_c$ and $\bf 8_s$ representations.

\begin{table}
\caption{The sixteen states of the representation space of
the SO(8) Clifford algebra corresponding to the quantization in the
$|n_m|=1$ monopole sector are sorted according to their $S$ and
electric charges. Each of these states is actually a
complete $N=2$ hypermultiplet. The transformation properties
of the states under Spin(8) (for $m=0$)
and ${\mathrm SU(2)_{v}}\times {\mathrm SU(2)_{c}}\times
{\mathrm SU(2)_{s}}$ (when $m\not = 0$) are also indicated.
$S_0$ is the a priori unknown $S$ charge of the state $\zero $.}
\begin{tabular}{|p{7pc}|p{3pc}|p{14.5pc}|p{2.45pc}|p{3pc}|}
\hline
${\mathrm SU(2)}_{v}\times {\mathrm SU(2)}_{c}\times {\mathrm SU(2)}_{s}$ 
rep. 
& Spin(8) rep.& States &
$n_e$ & $S$ \\
\hline\hline
$({\bf 1},{\bf 2},{\bf 1})$ & $\bf 8_c$ & $\psi _3\zero $, $\psi _4\zero $
& odd & $S_0$ \\
\hline
$({\bf 1},{\bf 1},{\bf 2})$ & $\bf 8_s$ & $\zero $, $\psi _3\psi _4\zero $
& even & $S_0$ \\
\hline\hline
$({\bf 2},{\bf 1},{\bf 2})$ & $\bf 8_c$ & $\psi _1\zero $, $\psi _2\zero $,
$\psi _1\psi _3\psi _4\zero $, $\psi _2\psi _3\psi _4\zero $
& odd & $S_0+1$ \\
\hline
$({\bf 2},{\bf 2},{\bf 1})$ & $\bf 8_s$ & $\psi _1\psi _3\zero $,
$\psi _1\psi _4\zero $, $\psi _2\psi _3\zero $, $\psi _2\psi _4\zero $
& even & $S_0+1$ \\
\hline\hline
$({\bf 1},{\bf 2},{\bf 1})$ & $\bf 8_c$ & $\psi _1\psi _2\psi _3\zero $,
$\psi _1\psi _2\psi _4\zero $
& odd & $S_0+2$ \\
\hline
$({\bf 1},{\bf 1},{\bf 2})$ & $\bf 8_s$ & $\psi _1\psi _2\zero $,
$\psi _1\psi _2\psi _3\psi _4\zero $
& even & $S_0+2$ \\
\hline
\end{tabular}
\end{table}
\subsection{Subtleties with the $S$ charge}
In the BPS mass formula (\ref{BPSmass},\ref{cc}) appears, in addition to the
electric and magnetic quantum numbers $n_e$ and $n_m$, a term 
$ms/\sqrt{2}$ proportional to the mass. The computation of $s$ for each BPS
state is crucial in order to have quantitative predictions
for the physical masses, as we will need in the following.
However, the meaning of the constant
$s$ is not clear at first sight. It cannot be identified with the
physical $S$ charge, because the latter has a non trivial dependence
on the Higgs expectation value due to the spontaneous breaking of
CP invariance. This is, for the $S$ charge, a similar effect as the
Witten phenomenon for the electric charge
\cite{WITeff}, and was discussed in \cite{FER}. There it was shown that
the non trivial part of the physical $S$ charge is actually
automatically included in the periods $a$ and $a_D$, and is responsible
for the constant shifts these variables can undergo under duality
transformations. This possibility is related to the fact that
$a$ and $a_D$ are period integrals of a meromorphic one form having
poles with non zero residus \cite{SWII}, see Appendix A. The constants $s$ are
thus simply a constant part of $S$ not already included in $a_D$ and $a$.
They can always be determined by consistency considerations, see
Section 4.4 and the discussion in Section 5.
\section{The $\tau\rightarrow -1/\tau $ transformation}
In this Section, some arguments are presented which lead
to a clear physical understanding of {\em why}
states with any magnetic charges must exist
in the theories we are studying, in the context of quantum field theory
alone.

We will heavily rely on the Seiberg-Witten low energy effective
action, encoded in the elliptic curve presented in \cite{SWII}.
Thus this Section begins with a brief review of this result,
emphasizing that it is by now established at a high level
of rigour, and above all {\em does not} rely on any duality assumption. 
This short survey will also serve to set our notations.
\subsection{The Seiberg-Witten curve}
Because of $N=2$ supersymmetry, the variables $a_{D}$ and $a$, which not 
only completely determine the form of the low energy effective action 
(up to two derivatives and four fermion terms), but also enter the
BPS mass formula (\ref{BPSmass}), must be analytic functions of the
gauge invariant coordinate $z$ on the Coulomb branch \cite{SWII}. These
analytic functions have branch cuts and SL$(2,\Z )$ monodromies due to
singularities caused by charged particles becoming massless for some 
values of $z$.
The number and type of singularities occuring on the Coulomb branch can 
most easily be found by analyzing the theory in a limit where the mass
$m$ is very large compared to some scale
$\Lambda $ which represents the dynamically generated scale of the 
asymptotically free theory we obtain after integrating out the 
supermassive hyper(s). Quantitatively, this limit corresponds 
to\footnote{The numerical factors depend on the regularization 
scheme \cite{pouliot}. We choose here the same conventions as in 
\cite{SWII}.}
\begin{equation}
\label{RG}
4m^4 q(\tau ) = \Lambda ^4,\quad m\rightarrow\infty ,\quad q\rightarrow 0,
\end{equation}
with
\begin{equation}
q(\tau )=e^{2i\pi\tau }.
\end{equation}
Let us focus for concreteness on the SO(3) theory.
In this regime, and at scales of order $\Lambda $, the theory must look like
the pure gauge theory, since the ultra-massive quarks must decouple.
The singularity structure of the pure gauge theory was already greatly
deduced in \cite{SWI}, and it was recently studied at an 
even higher degree of rigour \cite{SWIrig}. The result is that, 
without making any duality assumption, one can prove that we 
have two singularities at strong coupling whose
monodromy matrices are conjugate to $T^{2}$.
Moreover, we must 
have at scales of order $m$, where a tree level analysis must be valid
when $q\rightarrow 0$, an additional singularity coming from an 
elementary quark becoming massless.
The monodromy matrix is here again 
conjugate to $T^{2}$, as can be deduced with the help of the $\beta $ 
function of the low energy theory \cite{SWII}, or even directly in the 
microscopic theory \cite{FER}. Actually, the monodromy transformation
for $(a_{D},a)$ around a singularity at $z_{0}$ due to a state 
$(n_{e},n_{m})_{s}$ becoming massless is
\begin{equation}
\label{mono}
\pmatrix{a_{D}\cr a\cr }(e^{2i\pi}z_{0})=M_{(n_{e},n_{m})}
\pmatrix{a_{D}\cr a\cr }(z_{0})-{2ms\over\sqrt{2}}\,\pmatrix{n_{e}\cr
n_{m}\cr},
\end{equation}
where
\begin{equation}
\label{monomat}
M_{(n_{e},n_{m})}=\pmatrix{1-2n_{e}n_{m} & 2n_{e}^{2}\cr
-2n_{m}^{2} & 1+2n_{e}n_{m}\cr }.
\end{equation}

Searching for an elliptic curve of the form $Y^{2}=P(X)$, where $P$ 
is a polynomial of degree three, whose moduli space will reproduce the 
singularity structure described above, is then a fairly simple mathematical 
exercice. The solution is unique and given by \cite{SWII}
\begin{equation}
\label{curve}
Y^{2}=\prod_{j=1}^{3}\Bigl(X-E_{j}(\tau )z-{1\over 4}E_{j}^{2}(\tau ) 
m^{2}\Bigr),
\end{equation}
where
\begin{equation}
E_1(\tau )={1\over 3}(\theta _2^4 + \theta _3^4),\quad
E_2(\tau )=-{1\over 3}(\theta _1^4 + \theta _3^4),\quad
E_3(\tau )={1\over 3}(\theta _1^4 -\theta _2^4)
\end{equation}
and 

\begin{equation}
\label{thetadef}
\theta _1 (\tau ) = \sum _{n=-\infty}^{\infty} 
q^{(n+1/2)^2/2}, \quad
\theta _2 (\tau ) = \sum _{n=-\infty}^{\infty} (-1)^n q^{n^2/2},\quad
\theta _3 (\tau ) = \sum _{n=-\infty}^{\infty} q^{n^2/2}.
\end{equation}

The discussion above was for the SO(3) theory, but as noted in 
\cite{SWII}, the singularity structure and thus the curve is exactly 
the same in the $N_{f}=4$ theory, provided one chooses bare masses  
$m_{1}=m_{2}=m/2$ and $m_{3}=m_{4}=0$ for the matter hypermultiplets,
uses the different sets of convention introduced in Section 
2.1, and expresses the curve in terms of a dimensionless constant $\tau $
which is $\tilde\tau =\theta /\pi + 8i\pi /g^{2}$ only at the tree 
level but receives one loop as well as non perturbative corrections
\cite{instNf4}. Though this useful relation
will allow us to study the SO(3) and the $N_{f}=4$ theories
in parallel, be careful that it is only a formal one.
For instance, only even, $2n$-instantons corrections exist in the $N_{f}=4$ 
theory, whereas all $n$-instantons contribute in the $SO(3)$ theory. 
With our conventions, these corrections will both be proportional to 
$q^{n}$. There are also important differences at the level of the 
spectra. One of them is that in the 
SO(3) theory, a state $(1,0)$ may correspond to a vector multiplet (W 
bosons), whereas in the $N_{f}=4$ theory the Ws are labeled as 
$(2,0)$. Moreover, due to the flavour symmetry, the singularities
in the $N_f=4$ theory are produced by SU(2) doublets of hypermultiplets
becoming massless. The general form of the monodromy matrices
(\ref{monomat}) remains valid since if the solution for the periods is noted
$(a_D,a)$ for the SO(3) theory, it will be $(a_D/2,a/2)$ for the
$N_f=4$ theory.

The parameter $z$ appearing in (\ref{curve})
is a good gauge invariant coordinate on the Coulomb
branch, which is related to the physical expectation value
$\langle {\mathrm tr }\, \phi ^2\rangle $ by a formula of the form
\begin{equation}
z=\langle {\mathrm tr }\, \phi ^2\rangle - {1\over 8} m^2 E_1(\tau ) +
m^2 \sum_{n=1}^{\infty}c_{n}q^{n}.
\end{equation}
The $c_{n}$ correspond to instanton corrections \cite{instNf4,dorbis}.
We conventionally extracted
the combination  $\langle {\mathrm tr }\, \phi ^2\rangle-
m^2 E_1(\tau )/8$ because it tends towards the physical parameter
$\langle {\mathrm tr }\,\phi ^2\rangle $ under the renormalization group 
flow towards the pure gauge theory or the $N_{f}=2$ massless theory 
(\ref{RG}).

The transformation properties of the parameters 
$(z,\tau ,m)$ of the theories under 
S duality can easily be read from the curve (\ref{curve}). Under a general 
SL$(2,\Z )$ transformation given by a matrix
\begin{equation}
D=\pmatrix{a&b\cr c&d\cr },\quad ad-bc=1,
\end{equation}
we have
\begin{equation}
\label{dualt}
\tau \rightarrow \tau ^D={a\tau +b\over c\tau +d}\raise 2pt\hbox{,}\quad
z \rightarrow  z^D=(c\tau +d)^{2}z,\quad
m \rightarrow m^D=m.
\end{equation}
The fact that these strong-weak coupling transformations can be 
coherently implemented on the curve (\ref{curve}) is already a non-trivial 
evidence that the theories may be self-dual \cite{SWII}.
Below, we will investigate the transformation 
properties of the variables $a_{D}$ and $a$ (or equivalently of the 
electric $n_{e}$ and magnetic $n_{m}$ charges). This will give rise 
to predictions on the BPS spectra. Checking that these predictions are 
indeed true will be the main goal of the forecoming Sections.
\subsection{The analytic structure and the duality transformations of the 
periods}
In the remainder of Section 3, we will set the bare $\theta $ angle to 
zero. The 
very interesting physics associated with the possibility of having a 
varying real part in $\tau $ will be considered in Section 4.

To guess what the analytic structure (position of the singularities 
and of the branch cuts) looks like, it is simplest to choose the 
parameters so that we are not far away from the pure gauge theory
(or massless $N_f=2$), where 
we easily control what is going on. At strong coupling, 
we have two singularities on the real axis, at 
points $z_{2}$ and $z_{3}$ such that $z_{3}-z_{2}\simeq 2\Lambda ^{2}$,
with an analytic structure already worked out in \cite{FB,BF}. 
Moreover, at weak coupling, we have a quark becoming massless at 
$z_{1}$, with the following asymptotics for $a_{D}$ and $a$:
\begin{eqnarray}
\label{asympq}
a_{D}&\simeq &a_{D}(z_{1})-{i\over\pi }\,
\Bigl( a-{m\over\sqrt{2}}\Bigr)\ln\Bigl(a-{m\over\sqrt{2}}\Bigr),
\nonumber\\
a&\simeq &{m\over\sqrt{2}}\cdotp\\
\nonumber
\end{eqnarray}
The exact location of the singularities $z_{j}$ can be found by setting the 
discriminant of the curve (\ref{curve}) to zero,
\begin{equation}
\label{singpos}
z_{j}={1\over 4}E_{j}(\tau ) m^{2}.
\end{equation}
\begin{figure}
\label{figure1}
\epsfxsize=14.7cm
\epsfbox{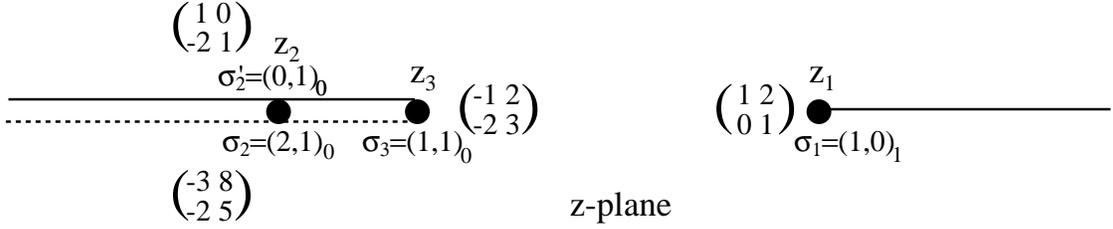}
\caption{The analytic structure of the solution for $(a_{D},a)$. The 
singulatities at $z_j$
are represented by bullets, the solid lines correspond to the 
branch cuts of $a_{D}$ and the dotted line to the branch cuts of $a$. The 
quantum numbers $(n_{e},n_{m})_{s}$ of the particles becoming massless at 
the singularities, as well as the corresponding monodromy matrices,
are also indicated.}
\end{figure}
We have $z_{1}>z_{3}>z_{2}$.
We choose the 
position of the branch cut for the logarithm appearing in the
asymptotics (\ref{asympq}) in such a way that it does not disturb
the analytic structure of the pure gauge theory. This is the most
natural choice, since the existence of a singularity at
a scale $m>>\Lambda $ cannot influence the physics at scale
$\Lambda $. In Figure 1 are also 
indicated the quantum numbers of the particles becoming massless.
Note that due to a ``democracy'' between the dyons \cite{SWI,FB,BF},
the most general choices for the electric and magnetic charges of these
states are
\begin{equation}
\label{sigma}
\sigma _1 =(1,0),\enspace\sigma _2 =(n_e+1,1),\enspace\sigma _2 ' =
(n_e-1,1),\enspace\sigma _3 = (n_e,1),
\end{equation}
where $\sigma _j$ corresponds to the singularity $z_j$ and $n_e$ is any
integer. Our choice is $n_e=1$.
Note also that due to the cuts, the particle becoming massless at $z_{2}$ is 
described by two different sets of integers, $\sigma _2$ or $\sigma _2'$,
depending on whether one looks 
from the $\IM z<0$ or $\IM z>0$ half-plane. This shows that one must 
introduce two different monodromy matrices at $z_{2}$, 
\begin{equation}
\label{matrices}
M_2=M_{(2,1)}=\pmatrix{-3 & 8\cr -2&5\cr},\enspace {\mathrm and}\enspace
M_2'=M_{(0,1)}=\pmatrix{\hfill 1&0\cr -2&1\cr}.
\end{equation}
The other monodromy matrices are
\begin{equation}
\label{matrices2}
M_1=M_{(1,0)}=\pmatrix{1&2\cr 0&1\cr},\quad
M_3=M_{(1,1)}=\pmatrix{-1&2\cr -2&3\cr}.
\end{equation}
It is now possible to study the transformation properties 
of $(a_D,a)$ (or equivalently of $(n_e,n_m)_s$) under duality.
The naive guess would be that $(a_{D},a)$ transforms 
exactly according to the corresponding SL$(2,\Z )$ matrix $D$
acting on the parameters of the microscopic theory as in (\ref{dualt}).
However, we will see that this is not true in general. We must 
introduce an SL$(2,\Z )$ matrix $D_{\mathrm eff}$ which acts on the 
low energy effective action (i.e. $a_D$ and $a$)
while $D$ acts on the microscopic parameters.
This subtlelty stems from the fact that the monodromy group
associated with the curve (\ref{curve}) is
$\Gamma (2)$, the subgroup of SL$(2,\Z )$ consisting in the matrices 
congruent to the identity modulo 2.
The only straightforward and correct statement that can be done is 
thus that $D_{\mathrm eff}=D$ mod 2.
Moreover, in the massive theories, $a_D$ and $a$ can also pick up some
constants under a duality transformation. This is allowed by the form
of the BPS mass formula (\ref{BPSmass},\ref{cc}), since a shift in
$a_D$ or (and) $a$ can be reabsorbed in a shift of $s$. A very similar
and related phenomenon occurs when one is studying the duality transformations
of the low energy theory \cite{SWII,FER}, but one must keep in mind that
by now we are studying duality transformations of the whole, microscopic,
theory.

Let us be more concrete. What we want to deduce is a formula
relating the dual theory solution characterised by the parameters
$(z^D,\tau ^D,m^D)$ given by (\ref{dualt}) and whose periods are
\begin{eqnarray}
\label{dualperiods}
a_D^{D}(z,\tau ,m)& =& a_D(z^D,\tau ^D,m^D),\nonumber\\
a^D(z,\tau ,m)& =& a(z^D,\tau ^D,m^D),\\
\nonumber
\end{eqnarray}
to the original solution described by $(a_D(z,\tau ,m),a(z,\tau ,m))$.
Let us focuss in this Section on the transformation $D=S$.
The analytic structure of $(a_D^D,a^D)$ can be readily deduced from the one
of $(a_D,a)$ depicted in Figure 1.
Indeed, as $E_{1}(-1/\tau )=\tau ^{2}E_{2}(\tau )$,
$E_{2}(-1/\tau )=\tau ^{2}E_{1}(\tau )$ and $E_{3}(-1/\tau )=
\tau ^{2}E_{3}(\tau )$, the effect
of the duality transformation will simply be to exchange the 
singularities at $z_{1}$ and $z_{2}$ while the singularity at $z_{3}$
remains fixed. The dual theory thus 
has singularities at exactly the same points as the original
theory (a direct consequence of the self-duality of the curve
(\ref{curve})), but the quantum numbers of the states becoming massless
at a given point, as well as the position of the
branch cuts, are changed, see Figure 2.
\begin{figure}
\label{figure2}
\epsfxsize=14.7cm
\epsfbox{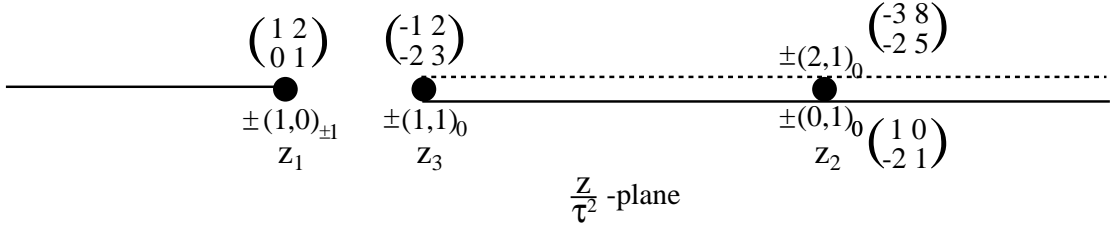}
\caption{The analytic structure of the dual theory. It should be 
compared with the one of the original theory depicted in Figure 1.}
\end{figure}
This is enough to deduce that
\begin{eqnarray}
\label{periodstrans}
{\mathrm for}\enspace\IM z>0&\,:\,&
\left\lbrace
\begin{array}{l}
\displaystyle
a_D(z,-1/\tau ,m)= a(z/\tau ^2,\tau ,m) -2a_D(z/\tau ^2,\tau ,m) + 
{m\over\sqrt{2}}\\
\displaystyle
a(z,-1/\tau ,m)= -a_D(z/\tau ^2,\tau ,m) + {m\over\sqrt{2}}\\
\end{array}
\right.\nonumber\\
{\mathrm and\ for}\enspace\IM z<0&\,:\,&
\left\lbrace
\begin{array}{l}
\displaystyle
a_D(z,-1/\tau ,m)=-a(z/\tau ^2,\tau ,m)+ {m\over\sqrt{2}}\\
\displaystyle
a(z,-1/\tau ,m)= a_D(z/\tau ^2,\tau ,m)-2a(z/\tau ^2,\tau ,m)+
{m\over\sqrt{2}} \cdotp\\
\end{array}
\right.\nonumber\\
\end{eqnarray}
The shifts by $m/\sqrt{2}$ come from the fact
that a dyon is exchanged with
a quark, and these two states have different $s$. Moreover,
the S duality transformation does not simply exchange $a$ and $a_D$
(up to a sign, and up to the shifts).
This is not surprising since
the state becoming massless at $z_3$ is $(1,1)$ in both formulations,
and this state is not self-dual (is not an eigenvector of S).
Second, to eliminate the cut between $z_3$ and $z_1$ in the dual
formulation (and thus recover the same structure for the cuts as in the
original formulation), it is necessary to perform an SL$(2,\Z )$
transformation, given on one side of the cut by the monodromy 
$M_{(0,1)}$. This also
explains why the transformation law depends on the sign of $\IM z$. 
\subsection{Implications of S duality on the dyon spectrum}
We are by now in a position to state precisely what the duality
$\tau\rightarrow -1/\tau $ implies for the
dyon spectrum. These predictions will be tested in the following.

At a very general level, suppose you postulate that the theories
described by the variables $(a_D,a)$ and $(\tilde a_D,\tilde a)$ are
equivalent. If
\begin{eqnarray}
\label{galtrans1}
\pmatrix{\tilde a_D\cr\tilde a\cr}=M\pmatrix{a_D\cr a\cr} +
{m\over\sqrt{2}}\,\Sigma ' &\quad\Longleftrightarrow\quad &
\pmatrix{a_D\cr a\cr}=M^{-1}\pmatrix{\tilde a_D\cr\tilde a\cr}
-{m\over\sqrt{2}}\,\Sigma \nonumber \\ \Sigma ' &=& M\Sigma ,\\
\nonumber
\end{eqnarray}
where $M$ is an SL$(2,\Z )$ matrix and
$\Sigma $ and $\Sigma '$ are $\Z ^2$-valued constant sections
(this is the most general transformation law we will encounter),
then the existence of a state $\sigma =(n_e,n_m)$ of charge $s$ in the
theory $(a_D,a)$ will imply the existence of a state $\sigma '=
(n_e',n_m')$ of charge $s'$ in the theory $(\tilde a_D,\tilde a)$ with
\begin{equation}
\label{galtrans2}
\sigma '=M\sigma ,\enspace s'=s+\sigma\cdot\Sigma =s+\sigma '\cdot
\Sigma '.
\end{equation}  
The $\cdot$ product denotes the standard symplectic product,
\begin{equation}
\label{symplecprod}
(p,q)\cdot (r,s)=ps-qr.
\end{equation}
The formula (\ref{galtrans2}) is a straighforward consequence of the
BPS mass formula (\ref{BPSmass}).

Let us apply (\ref{galtrans2}) to the S transformation $\tau\rightarrow
-1/\tau $ described by (\ref{periodstrans}).
Suppose that you have a state $(n_e,n_m)_s$ in the theory at given
$z$, $\tau $ and $m$. This state may be a vector or a
hyper multiplet of $N=2$ supersymmetry, and lie in a given representation
of the flavour group. Then S duality gives a prediction for the
spectrum of the theory whose parameters are
$(\tau ^2 z,-1/\tau ,m)$. If $\IM z>0$,
we must have there a state $(-2n_e + n_m,-n_e)_{s+n_m-n_e}$, in
a similar $N=2$ multiplet as the original state. The flavour
quantum numbers may be changed as discussed in Section 2.2.
If $\IM z<0$, the state predicted by duality will be
$(-n_m,n_e-2n_m)_{s+n_m-n_e}$. 

It is very important to realize
that these predictions of duality relate theories having different
parameters, even at the self-dual point $\tau = -1/\tau $. Thus a priori
they say nothing about the dyon spectrum of a given
theory (i.e., at fixed $z$, $\tau $ and $m$). One may be tempted
to argue that, due to the stability of the BPS states, the parameters
can be varied continuously without changing the spectrum. This means 
that a state which exists in the theory
$(\tau ^2 z,-1/\tau ,m)$ will also exist in the theory
$(z,\tau ,m)$. This reasoning is indeed correct when $m=0$,
but is wrong in the massive case, due to the
presence there of curves of marginal stability,
accross which otherwise stable
BPS states become degenerate in mass with multiparticle states. 
On these curves, the decay of BPS states is possible.
We will investigate these decays in the following.
Their mere existence implies that
{\em the self-duality of a general $N=2$ theory does not entail that  
the spectrum of BPS states is self-dual at a given point on the 
moduli space.} 
\subsection{A first grasp of the dyon spectrum}
Let us now penetrate at the heart of our problem: is there an easy
way to understand why the spectra of the massless theories 
should be self-dual? Can we readily imagine what the spectrum of the
massive theories look like?

There is a regime where the second question can be
easily answered, at least partially. By choosing the parameters as indicated
in (\ref{RG}), we can integrate out the quarks having a bare mass,
and we are thus left
with the pure gauge theory (for SO(3)) or with the
SU(2) theory with two massless flavours.
The spectra of these theories are well-known
\cite{FB,BF}. Discarding for the moment
the states having $s\not =0$, which can have
arbitrarily high physical masses 
when $m\rightarrow\infty $, two regions
in the Coulomb branch must be distinguished. Indeed, to the
possible decays of $s=0$ states into other $s=0$ states corresponds a
unique curve of marginal stability\footnote{We will see in Section 6 that
to the general decays of a dyon into states having arbitrary $s$ is
associated a whole family of curves.} ${\cal C}=\{\IM (a_D/a)=0\}$
which looks like an ellipse and contains the
singularities $z_2$ and $z_3$ \cite{courbe,FB}, see Figure 3.
\begin{figure}
\label{figure3}
\epsfxsize=14.7cm
\epsfbox{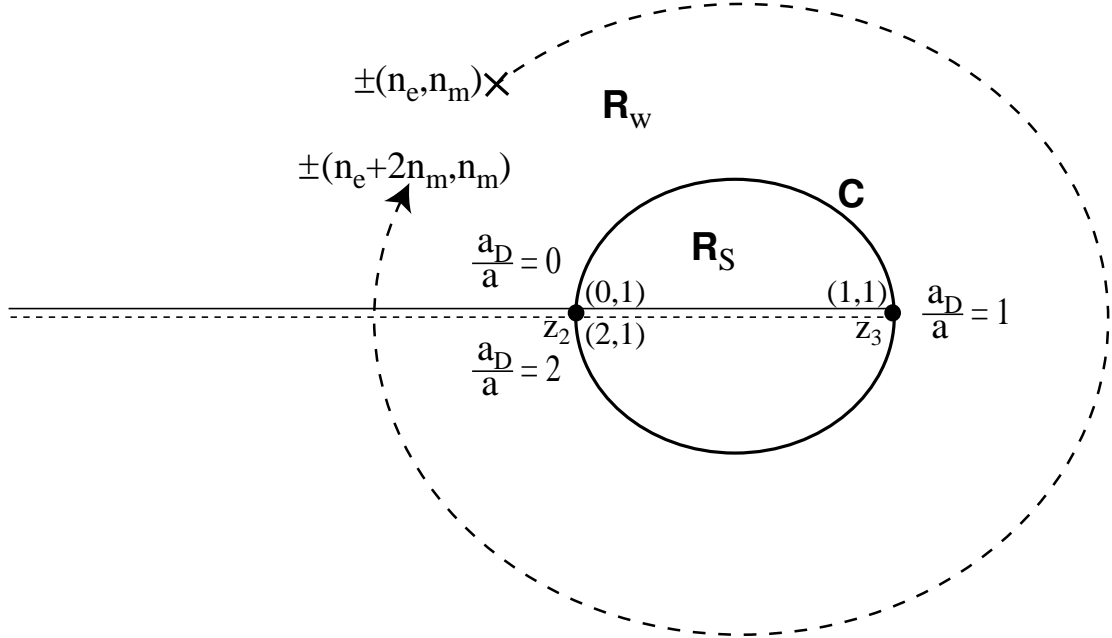}
\caption{The curve of marginal stability $\cal C$ for the pure gauge theory 
or for the $N_{f}=2$ massless theory.
The branch cuts as well as some values of the ratio $a_{D}/a$ are 
indicated. The path encircling the singularities used in the main text 
is also represented.}
\end{figure}
It can be shown that inside the curve (in the strong coupling region
${\cal R}_{\mathrm S}$) only the two BPS states which become massless
at $z_{2}$ or at $z_{3}$ can be present \cite{FB,BF}. These are $\pm 
(1,1)_{0}$ and $\pm (0,1)_{0}\equiv\pm (2,1)_{0}$ (for this latter 
state, two different descriptions in terms of electric and magnetic 
quantum numbers are necessary, due to 
the non trivial analytic structure \cite{FB,BF}).
Outside the curve, in what we
call the weak coupling region ${\cal R}_{W}$, the BPS spectrum 
can be understood in different ways. The most natural certainly is 
to perform a semiclassical analysis, which is valid here because of 
asymptotic freedom (in the context of the full, finite, 
theory, this means that we are considering a region 
of the Coulomb branch where
$\Lambda ^{2}\ll\langle{\mathrm tr}\,\phi ^{2}\rangle\ll m^{2}$).
This analysis first tells you that the 
perturbative states, created by the fundamental fields, must be present.
In the SO(3) theory, we have the W bosons 
$\pm (1,0)_{0}$ and the photon $(0,0)_{0}$
which lie in vector multiplets. In the $N_{f}=2$ theory 
we again have the W bosons, now represented as $\pm (2,0)_{0}$, the 
photon, but also 
the elementary quarks $\pm (1,0)_{1}$ and $\pm (1,0)_{-1}$.
Beyond this perturbative spectrum, we have a solitonic spectrum.
In the one monopole sector, we can construct all the 
states $(n_{e},1)$. The fact that 
all the integers $n_{e}$ are realized simply comes from the 
quantization of the periodic zero mode associated with electric 
charge rotations. Statements about higher magnetic charges would be 
much harder to do. It was shown in \cite{SZ} that no state with 
$n_{m}=2$ exists, but a complete analysis for $|n_{m}|\geq 3$ seems by now 
impossible, for the explicit form of the multimonopole metric is not 
known in general in these cases.
However, there is another method, initiated in \cite{FB},
which consists in remarking first that the spectum in ${\cal 
R}_{W}$ must be invariant under the transformation 
\begin{equation}
\label{scinv}
(n_{e},n_{m})\,\longrightarrow\, (n_{e}+2n_{m},n_{m}).
\end{equation}
This can be shown by 
transporting the state $(n_{e},n_{m})$ along a closed path encircling the 
two singularities and which does not cross $\cal C$, so that the state 
remains stable (see Figure 3). In this operation, 
we cross the cut $z<z_{2}$ and thus should perform an 
analytic continuation of the periods $a_{D}$ and $a$ in order to 
insure that the physical mass of the state varies continuously.
An analytic continuation $(\tilde a_{D},\tilde a)$ across a cut will 
in general be related to the main solution $(a_{D},a)$ by a relation 
of the form (\ref{galtrans1}) ($\Sigma '$ is indeed $\Z ^2$-valued
because the residues of the Seiberg-Witten differential form 
are $\pm m/\sqrt{2}$, see Appendix A). This shows that a state 
represented by $(n_{e},n_{m})_{s}$ on the one side of the cut will be
represented by $(\tilde n_{e},\tilde n_{m})_{\tilde s}$ on the other 
side, see (\ref{galtrans2}).
For the cut $z<z_{2}$, we will have $\Sigma =0$ and
$M=M_3M_2$ (or the inverse matrix depending on whether we 
turn clockwise or counterclockwise), which indeed generates the 
transformation (\ref{scinv}).
Since the states $(1,1)$ and $(0,1)$ 
must be present (they are responsible for the singularities), we 
immediately deduce that all the states $(n_{e},1)$ are also present, 
for any integer $n_{e}$. The reader might think that this is really a 
roundabout mean to show that electric charge is quantized.
However, this reasoning also provides a statement about the 
$|n_{m}|\geq 2$ sectors, powerful enough to show that no 
state with $|n_{m}|\geq 2$ can exist. Indeed, if such a state $(n_{e},n_{m})$
would exist, it would be associated with the whole tower of states 
$(n_{e}+2kn_{m},n_{m})$, for any integer $k$,\footnote{Note that
this has nothing to do with $\theta $ angle shifts by mean of which you
cannot show that all these states exist at the same time.}
and {\em one of these states 
will inevitably becomes massless} somewhere on the curve $\cal C$ since 
$a_{D}/a$ takes all real values between 0 and 2 on this curve \cite{FB}. 
This would produce an additional singularity on the Coulomb branch 
which does not exist.
 
The greatest virtue of this kind of 
reasoning is that one can generalize it to other cases.
We have just studied the RG flow (\ref{RG}) towards the pure gauge (or 
$N_{f}=2$) theory. It would be as natural to study the S dual of this 
RG flow, which would correspond to
\begin{equation}
\label{RGdual}
4m^{4}q^{D}(\tau )=\Lambda ^{4},\quad m\rightarrow\infty,\quad 
q^{D}\rightarrow 0,
\end{equation}
where
\begin{equation}
\label{RGdual2}
q^{D}(\tau )=e^{-2i\pi /\tau}.
\end{equation}
This is inherently a strong coupling limit where the semiclassical
method cannot be applied.\footnote{See, however, the remark below.}
The solution for the dual pure gauge (or massless $N_{f}=2$) theory 
is given by $(a_{D}^{D},a^{D})$, and its analytic structure has 
already been worked out, see Figure 2.
It is of course very tempting to try to repeat the analysis already 
made for the original theory. However, we cannot generalize
the reasoning at once because the position of the branch cuts are
now completely different. The fact that we will nevertheless obtain a 
spectrum in complete accordance with duality is a nice
evidence that S duality is indeed correct.
\subsection{Study of the dual pure gauge theory and of the dual 
$N_{f}=2$ massless theory}
The aim of this subsection is to work out the BPS spectra implied by 
the analytic structure depicted in Figure 2.
It is convenient to use, instead of 
$(a_{D}^{D},a^{D})$, the periods
\begin{equation}
\label{newsol}
\alpha _{D}=a_{D}^{D}-{m\over\sqrt{2}}\raise 2pt\hbox{,}\enspace
\alpha =a^{D}-{m\over\sqrt{2}}\cdotp
\end{equation}
This simply amounts to setting $s$ to zero for the two states $(1,0)$ 
and $(1,1)$ becoming massless.
We will note with an uppercase $D$ the $s$ charge computed with the 
new variables,
\begin{equation}
\label{news}
s^{D}=s+n_{m}-n_{e}.
\end{equation}
We have again a curve 
of marginal stability ${\cal C}^{D}=\{\IM (\alpha _D/\alpha )=0\}$,
which coincides with $\cal C$ because of (\ref{periodstrans}).
\begin{figure}
\label{figure4}
\epsfxsize=14.7cm
\epsfbox{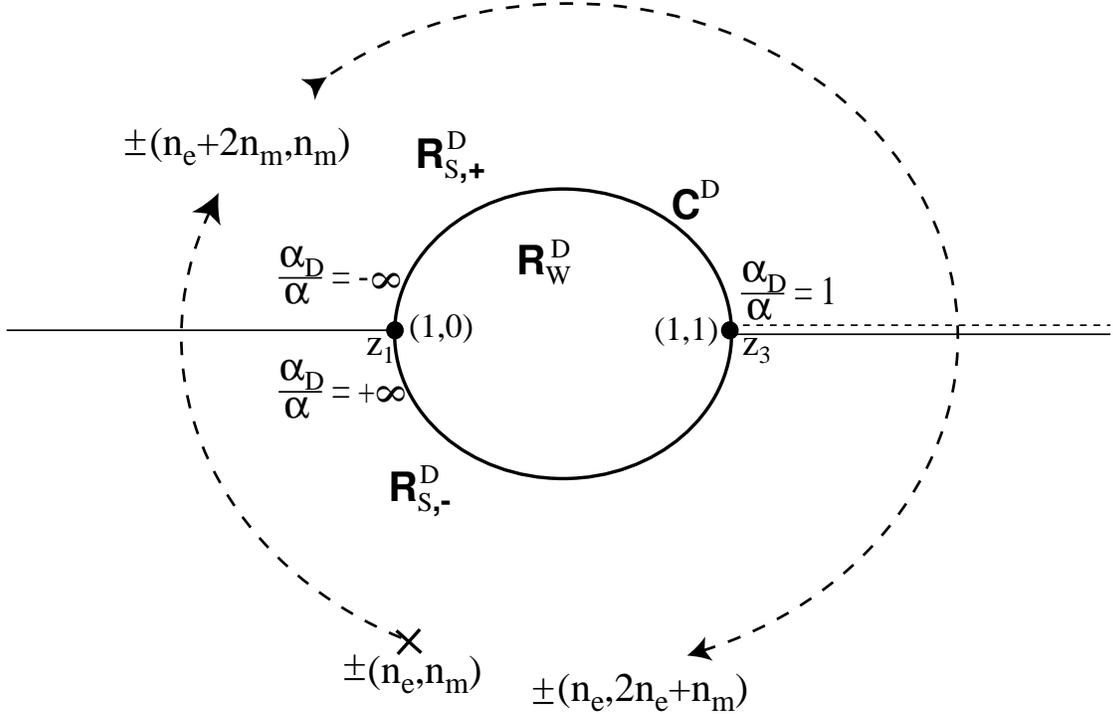}
\caption{The Coulomb branch of the dual pure gauge (or $N_f=2$)
theory obtained in the limit (\ref{RGdual}),
separated into the two regions ${\cal R}_W^D$ and
${\cal R}_S^D$ by the curve of marginal stability
${\cal C}^D$. The branch cuts, the
quantum numbers of the particles becoming massless, and the value
of $\alpha _D/\alpha $ at some points of ${\cal C}^D$ are indicated,
as well as the name of the different regions.
This configuration is the dual of the one displayed in Figure 3.}
\end{figure}
The Coulomb branch is thus separated into two regions. The one
inside the curve will be called in this context the weak coupling region
${\cal R}_W^D$ and is the dual of the strong coupling region 
${\cal R}_S$ one had previously. Outside the curve, we will have
the strong coupling region ${\cal R}_S^D$. 
These names deserve some comments. In any case, one should keep in mind that
the bare gauge coupling of the whole (finite) theory is
very large in the limit (\ref{RGdual}). Thus, by strong or weak coupling,
we only refer to the effective coupling which governs the low energy
physics. Moreover, even in ${\cal R}_W^D$ for instance, the effective theory
may be strongly coupled (it is near $z_3$ since there a magnetically 
charged particle is massless). The point is that the regions in 
${\cal R}_W^D$ where the coupling is strong can always be joined to
other regions in ${\cal R}_W^D$
where the coupling is weak, the dyon spectrum remaining unchanged. 
Indeed, the {\em electric} effective coupling is zero at $z_1$
where the quark is massless. This suggests that the BPS spectrum
might be probed in the vicinity of $z_1$ by using standard, 
``semiclassical,'' approximation schemes, since we do have in this
region a small parameter in the original, ``electric,'' theory.
Actually, if duality is correct (and we will prove that its
predictions indeed are concerning the dyon spectrum), the physics
should be the same in the weakly coupled region surrounding $z_1$
in Figure 4 as in the strongly coupled region surrounding
$z_2$ in Figure 3, and we should be able to account for
the ``strong coupling'' jumps  
of the BPS spectrum seen in \cite{FB,BF} by using such methods.
However, we will not try to do that here.

Let us rather apply our method to 
the weak coupling spectrum ${\cal S}_{W}^{D}$. It 
contains of course the states $(1,1)$ and $(1,0)$, which are 
responsible for the singularities and thus exist on both sides of the 
curve. We will now show that no other state, having $s^{D}=0$,
can exist. This can be understood by 
studying the variations of $\alpha _{D}/\alpha $ along the curve 
${\cal C}^{D}$. $\alpha _{D}/\alpha $ is related to $a_{D}/a$ 
by the formulas (\ref{newsol}) and (\ref{periodstrans}),
and $a_{D}/a$ varies monotonically from 0 to 2 
when one follows ${\cal C}^{D}$ clockwise starting from and ending at the 
singularity where $(1,0)$ is massless. This immediately implies that
$\alpha _{D}/\alpha $ takes all values between $-\infty $ and 1 in 
the upper half plane, and all values between 1 and $+\infty $ in the 
lower half plane. Thus any state $(n_{e},n_{m})$ existing in ${\cal 
R}_{W}^{D}$ will become massless somewhere on ${\cal C}^{D}$, at the 
point where $\alpha _{D}/\alpha =n_{e}/n_{m}$, and then 
must be either $(1,0)$ or $(1,1)$.
This is in perfect agreement with S duality.

When looking at the strong coupling spectrum, the complications come 
from the fact that one must consider analytic continuations across the 
cuts $z\geq z_{3}$ and $z\leq z_{1}$. We then need to introduce two 
different descriptions ${\cal S}^{D}_{S,+}$ and ${\cal S}^{D}_{S,-}$
in terms of electric and magnetic quantum 
numbers of the {\em same} physical spectrum ${\cal S}_{S}^{D}$, 
depending on whether we are in the upper or lower 
half plane. This is satisfactory from the 
point of view of duality, see Section 3.3.
Quantitatively, a state $(n_{e},n_{m})$ in ${\cal R}^{D}_{S,-}$ will 
be $M_1(n_{e},n_{m})=(n_{e}+2n_{m},n_{m})$ after crossing the 
cut $z\leq z_{1}$ and 
$M_3^{-1}(n_{e},n_{m})=(3n_{e}-2n_{m},2n_{e}-n_{m})$ after 
crossing the cut $z\geq z_{3}$. By looping around the curve of 
marginal stability, taking care of remaining in ${\cal R}^{D}_{S}$, 
we thus immediately deduce that ${\cal S}^{D}_{S,-}$ must be invariant 
under the subgroup of SL$(2,\Z )$ generated by $M_3 M_1$:
\begin{equation}
\label{invSDS}
{\cal S}^{D}_{S,-}=(M_3 M_1)^{p}{\cal S}^{D}_{S,-}=
(-1)^{p}\pmatrix{1&0\cr 2p&1\cr }{\cal S}^{D}_{S,-},\quad p\in\Z .
\end{equation}
Applying this transformation to the states $(1,0)$ and $(1,1)$, which 
we know must exist, will 
then generate states having {\em any} magnetic charge:
\begin{equation}
\label{provspec}
\{ \pm (1,p),p\in\Z \}\in {\cal S}_{S,-}^{D}\leftrightarrow
\{ \pm (1+2p,p),p\in\Z \}\in {\cal S}_{S,+}^{D}.
\end{equation}
Actually, this shows that all 
the states dual to the $(n_{e},1)$ states of the original theory will 
be present. Moreover, to each state $(n_{e},1)$ corresponds one and 
only one state in the dual theory, since there the states are in one to one 
correspondence with states becoming massless.
Are there any other states with $s^{D}=0$? If so, let $(n_{e},n_{m})$ 
be such a state in $\cal R_{S,-}^{D}$. If $n_{m}=0$, it is the 
state $(1,0)$ which indeed exists for it is responsible for one of 
the singularity.\footnote{No other state $(p,0)$, $s^D=0$ can exist since
they would modify the structure of the singularity at $z_1$.}
Otherwise, let us consider the ratio 
$r_{0}=n_{e}/n_{m}$. This ratio necessarily is in the interval
$]-\infty ,1[$, since otherwise the corresponding state would become 
massless somewhere on the curve ${\cal C}^{D}$ (note that we are 
working in the lower half plane and thus we see only half of the 
curve). As explained above, in the upper half plane the same state is 
described by other electric and
magnetic quantum numbers. Crossing the cut $z>z_{3}$, we will 
obtain another ratio $\tilde r_{0}=(3r_{0}-2)/(2r_{0}-1)$, which 
must be in 
the interval $]1,+\infty [$ unless the particle becomes massless 
somewhere on the Coulomb branch.
One can return in the lower half plane by crossing the cut $z<z_{1}$, 
and obtain again a new ratio $r_{1}$. By repeating this operation, we 
obtain two homographic sequences $r_{n}$ and $\tilde r_{n}$,
\begin{eqnarray}
r_{n} &=& {r_{0}\over 1-2r_{0}n}\nonumber\\
\tilde r_{n} &=& {(3+4n)r_{0} - 2\over (2+2n)r_{0} -1}\raise 2pt\hbox{,}\\
\nonumber
\end{eqnarray}
with the following important property: if, for some 
$n\in\Z $,\footnote{The negative values of $n$ are obtained by looping 
counterclockwise around ${\cal C}^{D}$.}
$r_{n}\in [1,+\infty [$ or $\tilde r_{n}\in ]-\infty ,1]$, then the 
state $(n_{e},n_{m})$ we started from must be one of the states already 
obtained above (see (\ref{provspec})). It is elementary to check that 
this is always the case, except if $r_{0}=0$. This case corresponds, 
in the SO(3) theory, to the dual of the W bosons, and is represented 
by $\pm (0,1)$ in ${\cal R}^{D}_{S,-}$ and by $\pm (2,1)$ in ${\cal 
R}^{D}_{S,+}$. In the $N_{f}=4$ theory, we would again have, in 
${\cal R}^{D}_{S,-}$, the state $\pm (0,1)$, 
now interpreted as being the dual 
of the elementary quarks, and also $\pm (0,2)$, interpreted as being 
the dual of the W.

Up to now, we did not take into account eventual decays of $s^D=0$
states into $s^D\not =0$ states along the path depicted in Figure 4,
although this path can cross 
curves of marginal stability corresponding to such decays.\footnote{For
a general discussion of curves of marginal stability, see Section 6.}
However, the states $s^{D}\not =0$ have a mass of order $m$, whereas
the states $s^{D}=0$ are likely to have a mass of order $\Lambda $ in
the region surrounding the singularities $z_{3}$ and $z_{1}$
we are considering. Since in
the RG flow, $m\rightarrow\infty $ while $\Lambda $ is fixed, the
decays are impossible and the curves of marginal stability simply
indicate that the inverse decay reaction from a $s^{D}\not =0$ to a
$s^{D}=0$ state is possible. One might put forward the fact that
states having $s^{D}\not =0$ but arbitrarily high electric and magnetic
charges could have a mass of order $\Lambda $ due to a subtle
compensation of the terms $n_{m}\alpha _{D}-n_{e}\alpha $ and
$ms^{D}/\sqrt{2}$ in the BPS mass formula. However such fine tuning,
which is indeed possible, can only occur
for some very special values of $n_{e}$ and $n_{m}$ and in some very 
special and small regions in the $z$-plane.
Nothing prevents us to avoid these special
points and to cross the curve of marginal stability nearby, where the
$s^{D}\not =0$ state again have a mass of order $m$.
I wish here to point out that, though we have considered the
possibility of these ``fine tuning'' points, they are very likely
to be unphysical. Indeed, if states producing these ``fine tuning''
points would exist, the theory would not have a well defined field theory
limit under the (dual) RG flow.

We have thus obtained the following results: states with any magnetic 
charges, dual to the dyons $n_{m}=1$, indeed exist, with the
correct multiplicities.
Thus, though the 
positions of the branch cuts are completely different in the original 
and in the dual theories, the deduced spectra are perfectly compatible
with duality. However, we have not yet proven that the states
corresponding to $r_0=0$ are really present; we simply know that they
may be. 
This point will be investigated in Appendix B.
\section{Self-dual field theories?}
\subsection{General discussion}
In the preceding Section, we have shown using simple arguments
that stable states 
having any magnetic charge, which are required by S duality, indeed 
exist. We were studying theories with non zero bare masses, but it 
is very tempting to think\footnote{See Section 7 for a rigorous proof}
that these states can be continuously 
deformed while remaining stable when $m\rightarrow 0$ and thus are at 
the origin of the existence of states having any magnetic charge in 
the massless theories as well. However, we will not obtain in this 
process all the states required by duality in the massless theories. 
For instance in the SO(3) ($N=4$) theory S duality implies the
existence of all the states $(n_{e},n_{m})$ with $n_{e}$ and $n_{m}$ 
relatively prime. Up to now, we only have the states $(q,n_{m})$
with $q=1$, and it is not difficult to realize in the
context of the semiclassical approach \cite{Segal}
that once you have one of these
states, then you will also have all the other states
$(1+kn_m,n_m)$, for any integer $k$. 
Actually, the missing states should come from configurations 
where states having any $n_e$ and $n_m$ relatively prime are massless
at the singularities. The fact that such configurations must exist
is clearly a necessary condition for the self-duality of the theories,
and does not follow straightforwardly from the SL$(2,\Z )$ invariance
of the curve (\ref{curve}). We saw above that for any imaginary value
of $\tau $ ($\theta =0$), the singularities are produced by states having
$|n_m|=0$ or $|n_m|=1$ alone. One must then study cases where
$\RE \tau\not =0$ and find a mechanism to understand how the
quantum numbers at the singularities can change. This is similar
to a phenomenon which is known to occur for instance in the massive SU(2)
asymptotically free theories, where semiclassical quarks must ``transmute''
into monopoles in order to account for the singularity structure
at strong coupling \cite{SWII}. 
However, there can be an important difference between this kind of
transmutation and the one we will see at work below. 
We will realize that two kinds of transmutations can occur.
In the first kind, the quantum numbers of the state are changed according
to a matrix of the form (8). We will see in the next
subsections that this is
the only kind of transmutation that can occur in the theories under
study in this paper, because singularities in the moduli space never
coincide, see (\ref{singpos}). 
In the second kind of transmutation, the state transforms
according to a matrix which is not of the form (8) (note that, in
any case, there are many of them, since the matrices (8)
are all conjugate to some power of $T$). 
These transmutations of the second kind 
are directly related to the necessity for two (or more)
singularities to coincide for some special values of the parameters
\cite{BFII}. These special points
are known to lead to superconformal theories \cite{superconf}.

Let us come back to our problem. We will soon
show that the quantum numbers
at the singularities can only be changed by matrices belonging to the
{\em monodromy} group $\cal G$ (in the cases under study, we will see that
any matrix of ${\cal G}=\Gamma (2)$ can be generated).
It is also clear that a necessary condition for the theories to be
self-dual is that any state, related to the original states becoming
massless by a {\em duality} group transformation, 
must become massless for some value of the coupling. If the $\sigma _j$ 
denote the electric and magnetic quantum numbers of the particles
becoming massless in the original configuration, we see that the
theory can be self-dual only if
\begin{equation}
\label{sdcond}
{\cal G} \{ \sigma _j\}={\cal D} \{ \sigma _j\},
\end{equation}
where ${\cal D}$ denotes the {\em duality} group.
For our purposes, ${\cal G}=\Gamma (2)/\{\pm 1\}$,
${\cal D}={\mathrm SL}(2,\Z )$,
$\{ \sigma _j \} =\{ (1,0);(2,1);(1,1)\}$ and ${\cal D}\{ \sigma _j\}=
\{ (p,q),p\wedge q=1 \}$. We thus have to check whether any
$(p,q)$, $p$ and $q$ relatively prime, can be generated by acting
with a $\Gamma (2)$ matrix on one of the three states
$(0,1)$, $(1,1)$ or $(2,1)$. This is indeed true because
SL$(2,\Z )/\Gamma (2)\sim S_3$ and that $S_3$
is generated by three transpositions whose corresponding matrices
(already described mod 2 in Section 2.2.2) can be chosen in order to
map a ``fundamental'' state, say $(1,0)$,
exactly into the three states $(1,0)$, $(1,1)$ and
$(2,1)$ we have at our disposal.
\subsection{The transformation $\tau\rightarrow\tau /(2\tau +1)$}
Let us start from a configuration, characterized by a certain
value of the bare coupling $\tau _0$ and of the bare mass $m$,
and whose analytic structure is of the same type as the one described
in Figure 1. We will allow the monodromy matrices $M_j$ to be of the
most general form, provided they satisfy the fundamental consistency
conditions
\begin{equation}
\label{consistency}
M_2' M_3 M_1 = M_1 M_3 M_2 = -1.
\end{equation}
To these monodromy matrices correspond states becoming massless,
characterized by their electric and magnetic quantum numbers
$\sigma _j =(p_j,q_j)$ and $s$-charge $s_j$. Note that
(\ref{consistency}) implies via (\ref{monomat}),
amongst other relations of the same type,
\begin{equation}
\label{consistency2}
\sigma _3 \cdot\sigma _2 = (\sigma _3 \cdot\sigma _1 )
                          (\sigma _1 \cdot\sigma _2 ).
\end{equation}
Actually, as we will only consider configurations which are
SL$(2,\Z )$ transforms of the original one depicted in Figure 1,
SL$(2,\Z )$ symplectic invariant relations can be checked or deduced using
$\sigma _1=(1,0)$, $\sigma _2 = (2,1)$, $\sigma _2' = (0,1)$ and
$\sigma _3 = (1,1)$. Some useful ones are
\begin{equation}
\label{relations}
\sigma _1 \cdot\sigma _3 =1,\enspace\sigma _2 =\sigma _3+\sigma _1,
\enspace\sigma _2'=\sigma _3 - \sigma _1.
\end{equation}
We must also have
\begin{equation}
\label{consistency3}
\sigma _2'=M_3 \sigma _2,\enspace M_2'=M_3M_2M_3^{-1},\enspace
s_2'=s_2 + 2s_3 \sigma _3 \cdot\sigma _2.
\end{equation}
Now, let us vary continuously $\tau $ from $\tau _0$ to $\tau _1=
\tau _0/(2\tau _0+1)$. During this operation, the singularities,
and with them the branch cuts, will move on the Coulomb branch,
as depicted in Figure 5.
\begin{figure}
\label{figure5}
\epsfxsize=14.7cm
\epsfbox{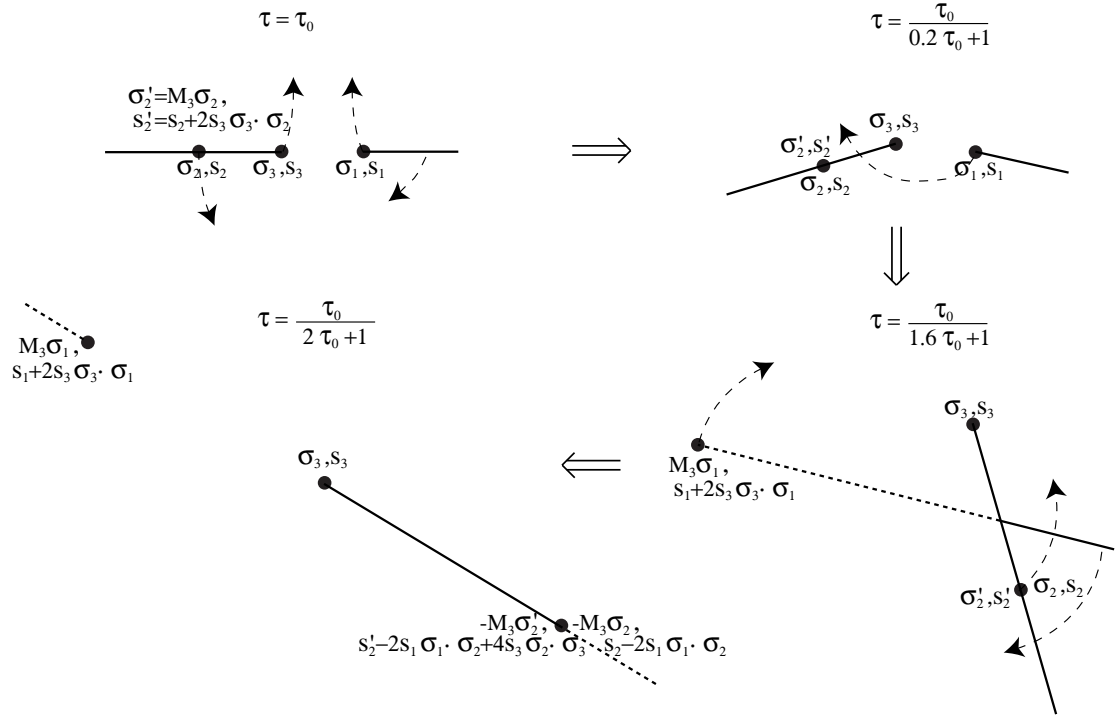}
\caption{Different steps in the analytic continuation from
$\tau =\tau _0$ to $\tau =\tau _1=\tau _0/(2\tau _0+1)$.
The picture is drawn for $\tau _0 =i$, $\tau =\tau _0/
(2\tau _0 t+1)$, $0\leq t\leq 1$, and $m$ real. 
The branch cuts are represented by solid or dotted lines depending
whether they lie on the original $\tau =\tau _0$ sheet or not.}
\end{figure}
Unavoidably, the singularity at $z_1$ will cross the cut
$[z_2,z_3]$; at this moment its quantum numbers are changed. According
to (\ref{mono},\ref{galtrans2}), we will have
\begin{equation}
\label{sigma1t}
\sigma _1\rightarrow M_3 \sigma _1,\enspace
M_1\rightarrow M_3 M_1 M_3^{-1},\enspace
s_1\rightarrow s_1 +2s_3\sigma _3 \cdot\sigma _1.
\end{equation}
The cut originating at $z_1$ will also sweep up one of the two singularities,
and then change the quantum numbers there.
We choose to change the singularity at $z_2$ (see Figure 5)
because it is the only case which leads to an analytic structure 
at $\tau =\tau _1$ which is perfectly similar to the one at $\tau _0$,
and this is convenient for our purposes.
The ambiguity associated with the position of the branch cut originating at
$z_1$ is further discussed in the next subsection.
As $\sigma _2$ sees the original cut produced by $\sigma _1$,
whereas $\sigma _2'$ sees the cut produced by $M_3 \sigma _1$, we will
have
\begin{eqnarray}
\label{sigma2t}
\sigma _2 &\rightarrow & M_1^{-1}\sigma _2,\enspace
M_2\rightarrow M_1^{-1}M_2 M_1,\enspace
s_2\rightarrow s_2-2s_1 \sigma _1\cdot\sigma _2 \nonumber\\
\sigma _2' &\rightarrow & M_3M_1^{-1}M_3^{-1}\sigma _2',\enspace
M_2'\rightarrow M_3M_1^{-1}M_3^{-1}M_2'M_3M_1M_3^{-1},\nonumber\\
&&\hskip 5cm
s_2'\rightarrow s_2'-2(s_1+2s_3\sigma _3 \cdot\sigma _1)M_3\sigma _1
\cdot\sigma _2'.\\
\nonumber
\end{eqnarray}
Finally, by using the consistency relations 
(\ref{consistency},\ref{consistency2},\ref{consistency3}), 
we see that at
$\tau =\tau _1 $ we are in essentially the same configuration as at
$\tau =\tau _0$, up to a $M_3$ transformation and to a rotation
and dilatation in the $z$ plane. Quantitatively we have
\begin{eqnarray}
\label{M3trans}
\pmatrix{a_D\cr a\cr}\Bigl( (2\tau +1)^2 z,{\tau\over 2\tau 
+1},m\Bigr)=
-M_3 & &\pmatrix{a_D\cr a\cr}\Bigl( z,\tau ,m\Bigr) + \nonumber\\
& &\qquad 2{m\over\sqrt{2}}\,
\Bigl( (s_1 + s_3)\sigma _3 -s_3 \sigma _1\Bigr).\\
\nonumber
\end{eqnarray}
Thus we have shown that we can change the quantum numbers of the
singularities with the matrix $M_3$ (the $s$ charges being also
changed according to (\ref{galtrans2})). 
\subsection{The transformation $\tau\rightarrow\tau +2$}
Performing $M_3$ transformations is not enough to prove
SL$(2,\Z )$ invariance. We need an additional transformation,
which can be found by studying the analytic continuation $\tau
\rightarrow\tau +2$ exactly along the lines of the preceding
subsection. We will not repeat this argument here, but we will rather
introduce another idea, which is maybe more heuristic, but has the 
merit of showing that the structure of the low energy effective 
action can imply the {\em equivalence} of theories related by some 
transformations of the monodromy group. We will elaborate more
on this kind of argument in Section 8.

The point is that by sweeping up the $z$ plane 
with the branch cut originating at $z_1$,
as shown in Figure 6, we can generate the transformation:\footnote{The
transformation
$\tau\rightarrow\tau +2$ would generate $M_3 M_2$ which is equivalent
to $M_1$ due to the relations (\ref{consistency}).}
\begin{equation}
\label{M1trans}
\pmatrix{a_D\cr a\cr}\longrightarrow M_1\pmatrix{a_D\cr a\cr}
-2\, {m\over\sqrt{2}}\, s_1 \sigma _1 .
\end{equation}
\begin{figure}
\label{figure6}
\epsfxsize=14.7cm
\epsfbox{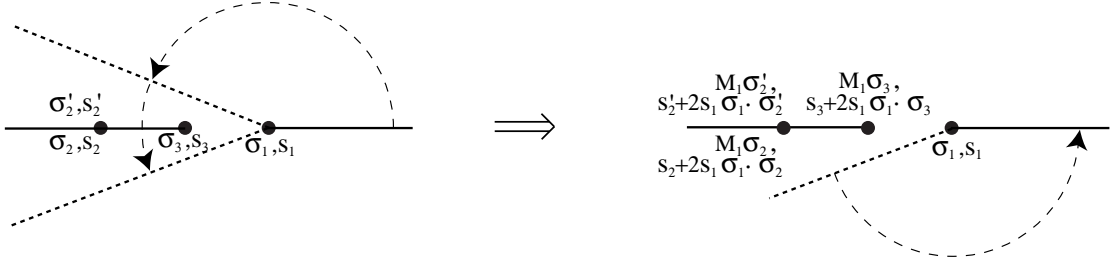}
\caption{By sweeping up the $z$ plane with the branch cut originating at
$z_1$, which is represented by dashed lines in intermediate positions,
we generate a $M_1$ transformation.}
\end{figure}
Doing this on the original configuration
of Figure 1 is innocent since it will generate a trivial 
$T^2$ transformation, which is interpreted as an unphysical  
relabelling of the electric quantum numbers $n_e$. 
It simply corresponds to a shift in the $\theta $ angle, which when
combined with the change in $n_e$ will leave the physical electric charge
and other physical observables unchanged. This shows that 
the position of the branch cut originating at $z_1$ is purely conventional
when $M_1=T^2$, and by continuity this ambiguity should be preserved
even when the $M_1$ matrix is general (we will see in the next 
subsection that any $M_1$ conjugated to $T^2$ by a 
$\Gamma (2)$ matrix can be generated by analytic continuation).
Note that we performed the transformation at {\em fixed} $z$, $\tau $
and $m$. In the original configuration where $M_1=T^2$,
this renders very explicit the fact that the transformation
$\theta\rightarrow\theta +4\pi $ is unphysical.\footnote{One might wish 
to recover that the transformation $\theta\rightarrow\theta +2\pi $
is unphysical. However, this is not that obvious in this 
framework, and we will limit ourselves to transformations of the
type of $T^2$.}
This also shows that the physical interpretation of the parameter
$\tau $ is no longer the same after analytic continuation: if we choose
to still use $\tau $ even after the $\theta $ angle shift, the real
part of $\tau $ clearly will no longer be $\theta /(2\pi )$ (or
$\theta /\pi$).
This is even more striking when $M_1$ is general. Then, the transformation
mixes the electric and magnetic charges in a non trivial way, still
at fixed $\tau $. This is not contradictory to the Dirac quantization
condition. Again, this shows that the duality transformation
associated with $M_1$ links together two theories which are
physically equivalent and thus can be labelled by the same parameters.
Of course, the physical interpretation of $\tau $ in the two
equivalent theories will not be the same. Similar remarks also apply
to $z$. To close this discussion,
note that the ambiguity in the interpretation of $\tau $ we
deal with in this subsection is related to the fact already
mentioned in subsection 3.2 that the SL$(2,\Z)$ matrix $D_{\mathrm eff}$
which acts on $(a_D,a)$ coincides with the
SL$(2,\Z )$ matrix $D$ which acts on $\tau $ and $z$, {\em only up to 
monodromy transformations.} This suggests that $\tau $ can only be 
defined modulo $\Gamma (2)$ and thus that the structure of the low 
energy effective action implies that the theory has at least an 
exact $\Gamma (2)$ duality symmetry. We will see in Section 8 that 
this can be understood on physical grounds, at least heuristically.
\subsection{Generating $\Gamma (2)$}
Let us come back to our original problem: can we generate any
$\Gamma (2)$ transformation (up to a sign)
by combining the two elementary transformations (\ref{M3trans},\ref{M1trans})
described in the previous subsections and
that we will denote abstractly $t_{3}$ and $t_{1}$?
Suppose that we execute the following sequence of transformations,
\begin{equation}
\label{seq1}
t_{3}^{q_{n}}t_{1}^{p_{n}}t_{3}^{q_{n-1}}t_{1}^{p_{n-1}}\ldots t_{3}^{q_{1}}
t_{1}^{p_{1}},
\end{equation}
starting from the configuration where the monodromy matrices are 
given by (\ref{matrices},\ref{matrices2}). One must take care of the fact 
that, after each step, the monodromy matrices are conjugated by the 
corresponding transformation. For instance, the transformation 
$t_{1}^{p_{1}}$ will indeed generate a $M_{1}^{p_{1}}$ transformation on the 
variables $a_{D}$ and $a$. However, the $t_{3}^{q_{1}}$ 
transformation that follows will be implemented by the matrix
$M_{1}^{p_{1}}M_{3}^{q_{1}}M_{1}^{-p_{1}}$, since the monodromy at
$z_{3}$ is no longer $M_{3}$ but $M_{1}^{p_{1}}M_{3}M_{1}^{-p_{1}}$.
The effect of conjugating the matrices at each step is
simply to reverse the order of the transformations in (\ref{seq1})
to which correspond, up to a sign, the SL$(2,\Z )$ matrix
\begin{equation}
\label{seq2}
M_{1}^{p_{1}}M_{3}^{q_{1}}M_{1}^{p_{2}}M_{3}^{q_{2}}\ldots
M_{1}^{p_{n}}M_{3}^{q_{n}}.
\end{equation}
Here $M_{1}$ and $M_{3}$ are fixed matrices given by 
(\ref{matrices},\ref{matrices2}).
We are thus able to obtain any matrix 
of the subgroup of SL$(2,\Z )$ generated by $M_{1}$ and $M_{3}$.
This is just $\Gamma (2)$ as we need to prove the self duality of the
spectrum, as explained at the end of Section 4.1. 

The $s$ charges of the $(n_e,n_m)$ states obtained above can also be
easily computed, by evaluating the product (\ref{seq2}) where the
monodromy matrices $M_1$ and $M_3$ are replaced by the corresponding
elements of the full monodromy group, which is the semi-direct product
$\Z ^2\times {\mathrm SL}(2,\Z)$ (see (\ref{M1trans}) and (\ref{M3trans})):
\begin{equation}
\label{seq3}
(-2\sigma _1,M_1)^{p_1}(2\sigma _3,-M_3)^{q_1}\ldots
(-2\sigma _1,M_1)^{p_n}(2\sigma _3,-M_3)^{q_n}.
\end{equation}
If
\begin{equation}
(n_e,n_m)=M_{1}^{p_{1}}(-M_{3})^{q_{1}}\ldots
M_{1}^{p_{n}}(-M_{3})^{q_{n}}\sigma _j ,
\end{equation}
then $s=n_e-n_m$ for $j=1$ or 3, $s=-1+n_e-n_m$ for $j=2$ and
$s=1+n_e-n_m$ for $j=2'$.
\subsection{Comparison with other approaches}
In order to put the present work into some perspective, I will
briefly discuss below some relations with other approaches.

Let me first explain how our line of reasoning can shed light on
some aspects of the semiclassical quantization. I will illustrate
this point on the following example.
It is well known in the context of the semiclassical approach 
\cite{SEN} that
the existence of dyon-monopole bound states with $|n_m|\geq 2$ in the
pure $N=2$ gauge theory would violate the self-duality of the
$N=4$ theory. This is due to the fact that such bound states are
in one to one correspondence with the cohomology classes of
{\em holomorphic} differentials on the reduced ($4n_m-4$ dimensional)
multimonopole moduli space in the case of the pure gauge theory,
while the differentials only need to be harmonic in the case
of $N=4$. This means that any $|n_m|\geq 2$
bound state of the pure gauge theory  corresponds
to $\em two$ bound states in the $N=4$ theory, the second state
corresponding to the antiholomorphic partner of the original
holomorphic differential form. This is in contradiction with $S$
duality which predicts the existence of one and only one bound state
of given $n_e$ and $n_m$.
From our point of view, this seemingly purely technical relation
between the two theories comes from the fact that they can be
joined together by continuously varying the bare mass and the
coupling and using the renormalization group flow. Stable states in the
pure gauge theory will then yield stable states in the
$N=4$ theory, by continuous deformation.\footnote{To rigorously prove
that the states remain stable during this process necessitates the
computation of the curves of marginal stability, see the next Sections.}
Moreover, we do not want any additional
$|n_m|\geq 2$ state in the $N=4$ theory
coming from the pure gauge theory, since we know that such states
originate from the existence of dual pure gauge theories configurations.

I wish now to point out one surprising aspect of our derivation,
and trace its profound origin in superstring dualities.
All our reasoning is based on a careful study of the low energy
effective action of the theories. Though it is well known that
the latter contains a lot of information about the {\em massive} states
(encoded in the BPS mass formula (\ref{BPSmass}))
in the context of $N=2$ supersymmetric theories, it might seem
astonishing that it also governs the {\em formation} of all the
stable bound states (coming in short multiplets).
Nevertheless, this fact, which appears for the first time in
\cite{FB}, is supported by increasing evidence. I hope that 
the present paper will convince the reader that it is both understandable
(if not natural) and quite general in the field theory framework.
However, a deeper understanding of this phenomenon comes
from string theory. There, in the framework of the heterotic-type II 
duality conjecture \cite{Hull},
you can argue that the BPS states of $N=2$ 
supersymmetric Yang-Mills theories correspond to a self-dual
non-critical string in six dimensions wrapped along geodesics
around the corresponding Seiberg-Wittern curve \cite{VafLer}.
The metric on the 
curve is directly related to a particular Seiberg-Witten 
differential, which can in principle be computed unambiguously.
As all these geometric data are contained in the low energy effective action, 
we understand why it completely determines the BPS 
spectrum.\footnote{Note
that a priori the Seiberg-Witten differential is only defined up to 
an exact meromorphic one-form. It is not known how to pick the right 
one form which will give the metric from the knowledge of the low 
energy effective action alone, though our analysis suggests that this 
might be possible.}
In the case of $N=4$, the 
low energy theory as well as the 
metric are trivial and the self-duality of the spectrum follows
easily \cite{Lowe} from the assumption that the type IIB string gives rise to
the $N=4$ gauge theory in some limit.
The cases of some SU(2) theories, including also very recently the SO(3)
theory studied in this paper, were addressed in
\cite{VafLer,Brand,Warner}.
The case of $N_f=4$ should also be within the stringy
approach capabilities.
Our results thus constitute non trivial tests of the string-string 
duality hypothesis.
\section{Uniqueness of the states}
We will mainly focus in this Section on the $N=4$ theory, in order to
avoid tiresome repetitions. In subsection 5.3, we will nevertheless briefly
point out the peculiarities of the $N_f=4$ theory, which do not
yield any new difficulties.
\subsection{Precise statement of the problem and useful remarks}
As there is only {\em one} multiplet in the Hilbert space of states
corresponding to the fundamental $N=4$ multiplet $(1,0)$, S duality
predicts the existence of only {\em one} $N=4$ multiplet $(p,q)$.
As indicated in the Table 1 of Section 2, such an $N=4$ multiplet
can be decomposed into three parts characterized by their physical
S charges $S_0$, $S_0+1$ and $S_0+2$.
When $m=0$, we can use CP invariance to show that
the existence of a state having $S$ implies the existence of another
state having $-S$, and that in a given monopole sector $n_m>0$,
$-n_m\leq S\leq n_m$.\footnote{Note that the $\theta $ angle
does not influence the physical $S$ charge \cite{FER}.}
Uniqueness of the $(p,q)$ state would therefore imply that $S_0=-1$, and
the three values of $S$ that should be realized correspond to
the CP invariant combination $\{ -1,0,1\}$.\footnote{This constraint
is directly related to the fact that the closed 
differential form corresponding to the $(p,q)$ state in the semiclassical
picture must be self-dual or anti self-dual.} The extremal values
$1$ and $-1$ correspond to states belonging to a $N=2$
hypermultiplet, while $S=0$ corresponds to a $N=2$ vector multiplet. 

The $(p,q)$ states we have generated in Section 4 in the massive
theory correspond to $N=2$ hypermultiplets, having a given charge
$S_H$. Because of $N=4$ supersymmetry in the $m\rightarrow 0$ limit,
these hypermultiplets will be associated to other states to form whole
$N=4$ multiplets $\{S_H,S_H+1,S_H+2\}$ or $\{S_H-2,S_H-1,S_H\}$.
The physical $S$ charges of a state $\sigma $ must be related
to the constants $s$ appearing in the central charge of the
supersymmetry algebra by a relation of the form \cite{FER}
\begin{equation}
\label{Sseq}
S=s+\sigma\cdot\Xi ,
\end{equation}
where $\Xi\in\Z ^2$. To determine $\Xi $, one can proceed as follows. 
When $m=0$, the solution for $a_{D}$ and $a$ is given by
\begin{equation}
\label{solm0}
a_{D}(z,\tau ,m=0)=(\tau +1-\epsilon _{z})\sqrt{2z},\quad
a(z,\tau ,m=0)=\sqrt{2z},
\end{equation}
where $\epsilon _{z}$ is the sign of $\IM z$. The (unphysical) cut 
for $a_{D}$ comes from the choice we did in Section 3 for the 
position of the branch cut originating at $z_{1}$ when $m\not =0$. 
When $m\not =0$, we will have
\begin{equation}
\label{asymequiv}
a_{D}\mathop{\sim}_{|z|\rightarrow\infty }
(\tau +1-\epsilon _{z})\sqrt{2z},\quad
a\mathop{\sim}_{|z|\rightarrow\infty } \sqrt{2z}.
\end{equation}
However, if $\Xi\not =0$, the physical $S$ charge contributes to 
$a_{D}$ and $a$, even at infinity, by an amount determined by $\Xi $ 
in such a way that
\begin{equation}
\label{limite}
{m\over\sqrt{2}}\, \Xi = \lim _{|z|\rightarrow\infty}
\left\lbrace\pmatrix{(\tau +1-\epsilon _{z})\sqrt{2z}\cr\sqrt{2z}\cr}
-\pmatrix{a_{D}(z,\tau ,m)\cr a(z,\tau ,m)\cr}\right\rbrace\cdotp
\end{equation}
$\Xi $ can then be computed straightforwardly by using the formulas 
for $a_{D}$ and $a$ given in Appendix A. One must note, however, that 
the sign of $\Xi $ remains arbitrary, since it can be changed by 
performing a Weyl gauge transformation. This is related to the fact 
that one can choose equivalently $(1,0)_{+1}$ or $(1,0)_{-1}$ to be 
massless at $z_{1}$ in the configuration of Figure 1. The convention 
we will choose hereafter is the following: $(1,0)_{+1}$ will be chosen 
to be massless at $z_{1}$ when $m>0$ and $(1,0)_{-1}$ when $m<0$ (we 
will only consider real values of $m$). These conventions will allow 
us to use the gauge invariance in a convenient way.
Finally, we have
\begin{equation}
\label{Sformula}
\Xi =(-\epsilon _{z}\epsilon _{m},0),\enspace
S=s+\epsilon _{m}\epsilon _{z}n_{m},
\end{equation}
where $\epsilon _{m}$ is the sign of $m$.
This formula implies in particular that the dyons found in Section 3 
by studying the dual RG flow (\ref{RGdual}) all have $S=\pm 1$ as 
expected. It is not
disturbing that $S$ may undergo a discontinuity from $\IM  z>0$
to $\IM z <0$. This discontinuity simply reflects the ambiguity
associated with CP invariance when $z$ or 
$\langle {\mathrm tr}\, \phi ^2\rangle $ are real.

To be concrete, we will hereafter consider a particular configuration, 
obtained from the original configuration depicted in Figure 1 by 
performing the transformation $t_{3}t_{1}t_{3}$ (see Section 4),
and which is described in Figure 7. This amounts to changing $a_{D}$ into 
$\tilde a_D=a_{D}-2a+\sqrt{2}\, m$ and $a$ into 
$\tilde a=4a_{D}-7a+2\sqrt{2}\, m$.
\begin{figure}
\label{figure7}
\epsfxsize=14.7cm
\epsfbox{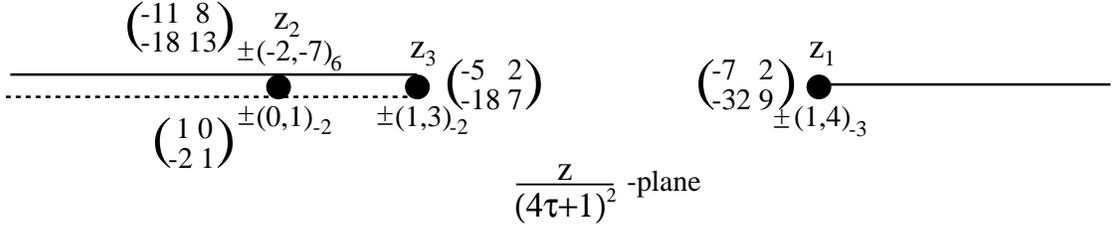}
\caption{The configuration obtained from the one depicted in Figure 1 
by performing the transformation $t_{3}t_{1}t_{3}$. With our 
conventions, the Figure corresponds to $m>0$.}
\end{figure}
This choice is completely arbitrary: {\em all our reasonings will be 
SL$(2,\Z )$ invariant and thus completely general.} However, 
we think that it is more eloquent to deal with states having magnetic 
charges three, four or seven than simply one,
for in this latter case unicity of 
the states is trivial from the semiclassical point of view.

We already know that the $N=4$ multiplet corresponding to a state
$(n_e,n_m)_{s}$ becoming massles is unique
since the singularities are produced by only one hypermultiplet.
Thus, what we need to show is that, for instance, the only 
allowed values of $s$ for a state $(1,4)$
should be $-3$ (charge of the hypermultiplet 
becoming massless) together with $-4$ and $-5$ (this would 
correspond to the case where the physical $S$ charge of $(1,4)_{-3}$ 
would be $+1$) or with $-2$ and $-1$ (when $S=-1$). 
\subsection{Uniqueness of the states}
Suppose that a state $(1,4)_s$ exists in the $N=4$ theory.
When a real mass $m$ is turned on, 
$(1,4)_s$ must produce a singularity at any point on the Coulomb
branch where it is massless, that is where $4\tilde a_D -\tilde a
+ms/\sqrt{2}=0$ which in the original variables reads
\begin{equation}
\label{uni1}
a=(4+s)\, {m\over\sqrt{2}}\cdotp
\end{equation}
To study this equation, at least in the large $|s|$ limit, one can use the
asymptotics (\ref{limite}) for $a$ at large $|z|$.
This suggest that (\ref{uni1}) might have solutions only when
$s\geq -4$ when $m>0$ and only when $s\leq -4$ when $m<0$ at points
\begin{equation}
\label{forbsing}
z_1^{(s)} (m)\simeq {1\over 4}(4+s)^2\, m^2.
\end{equation}
More rigorously, using (\ref{CPtrans}) and the monodromy $M_1$, one can 
show that $a(z)\in\R $ for $z\geq z_3$. Moreover, one can easily show
using the explicit formulas of $a$ and its derivative 
$\partial a/\partial z$ given in Appendix A that, in the case
$m>0$, $a$
is a monotonically increasing function for $z\in [z_3,+\infty [$,
with $0<a(0)<m/\sqrt{2}$ (this latter value is reached at $z_1=z_1^{(1)}$),
and $a\rightarrow\infty $ when $z\rightarrow\infty $.
Strictly similar results are true when $m<0$, which
proves that the simple analysis using the asymptotics is valid
except concerning $s=-4$, as one may have expected.
As the $z_{1}^{s}(m)$ exist for any $m>0$ when $s\geq -3$ and for 
any $m<0$ when $s\leq -5$, a singularity must be associated in the massive
theories to any state $(1,4)_s$ existing in the massless one,
provided $s\geq -3$ (if $m>0$) or $s\leq -5$ (if $m<0$). From this
we deduce that when $m>0$, which corresponds to the case depicted in
Figure 7, the state becoming massless at $z_1$, having
$s=-3$, must have a physical $S$ charge $+1$ (it is thus associated with
states having $s=-4$ and $s=-5$ when $m=0$), and that no other
$(1,4)_s$ state having $s>-3$ can exist when $m=0$. When $m<0$,
the state becoming massless at $z_1$ will have $s=-5$ and 
$S=-1$\footnote{This means that
the hypermultiplet component of the $N=4$ multiplet (see Table 1 in
Section 2) which becomes
massless is changed when $m$ changes sign with our conventions.}
and we see that no
other $(1,4)_s$ state having $s<-5$ can exist when $m=0$.
This proves the unicity.

Exactly the same reasoning can be applied to the states $(1,3)_s$.
The equation to study is in this case
\begin{equation}
\label{uni2}
a_D-a=(s+2)\, {m\over\sqrt{2}}\cdotp
\end{equation}
By using again (\ref{limite}), we can convince ourselves, and then
show rigorously, that when $\IM z>0$ (\ref{uni2}) will have solutions
when $s<-1$ (for $m>0$) or when $s>-3$ (for $m<0$). This is again
perfectly consistent with our previous analysis of the $s$ and $S$ charges.
When $m>0$, it is the state having $S=-1$ which is massless, and we 
showed that no state $(1,3)$ with $S<-1$ can exist; when $m<0$ it is the
state having $S=+1$ which is massless, and we showed that no state
with $S>1$ can exist. Thus the $N=4$ multiplet $(1,3)$ is indeed unique
and realizes the values $S=-1$, $S=0$ and $S=1$ as it should.

Finally, note that the case of the states $(2,7)_s$ and
$(0,1)_s$ can be easily handled for instance by using the previous results
and the $\tau\rightarrow
-1/\tau $ type duality transformation (\ref{periodstrans}), or by
repeating the same analysis as above.
Of course, the reasoning will not change whatever configuration
of the theory you may choose, and thus we have proven the unicity
of the states $(p,q)$ for any $p$ and $q$ relatively prime.
\subsection{The case of $N_{f}=4$}
In this theory, the solution is given by $a_D^{N_f=4}=a_D/2$ and
$a^{N_f=4}=a/2$, and the bare masses of the first and second flavours
are equal to $m/2$. The uniqueness of the $(p,q)$ states can then be
proven exactly as for the $N=4$ theory.
In particular, a $(p,q,S)$ hypermultiplet becoming massless when
$m\not =0$ will again be associated to two
other sets of states having physical charges $S-1$ and $S-2$, or
$S+1$ and $S+2$ when $m=0$, the Spin(8) flavour symmetry playing the
same r\^ole in this theory as $N=4$ supersymmetry in the previous
case. The main novelty is that 
one also has to study the uniqueness of the vector $(2p,2q)$ states.
Again one can repeat the arguments of the preceding subsection.
Equation (\ref{uni1}) will be replaced by
\begin{equation}
\label{uni3}
a={1\over 2}\,(4+s)\,{m\over\sqrt{2}}\cdotp
\end{equation}
From the analysis of the preceding subsection it follows that
this equation has solutions for any $s$ such that $|s-4|\geq 2$.
Moreover, by using the formula for $a$ given in Appendix A, one can 
show that $a(0)=|m/(2\sqrt{2})|$ when $\tau =i$ and thus the
equation (\ref{uni3}) will also have solutions for $|s-4|=1$ at least
for some values of $\tau $. Thus, the only $(2p,2q)$ states,
$(p,q)=(1,0)$ mod 2, that
might exist in the $m=0$ theory must have $S=0$,
as expected. The same reasoning works for the state $(2p,2q)$ such that
$(p,q)=(0,1)$ mod 2. However, when $(p,q)=(1,1)$ mod 2, one cannot
exclude directly the states having a physical $S$ charge $\pm 1$.
This seems to be a peculiarity of the vector particles: at fixed
$\tau $, these vector particles, if they existed, would not necessarily
lead to an unphysical singularity,
and hence their existence cannot be ruled out by 
this argument alone. Of course, this is not
really an obstacle for us. By changing $\tau $ to $\tau +1$ in the
original configuration, we exchange $z_2$ and $z_3$ and thus states
$(p,q)=(0,1)$ mod 2 with states $(p,q)=(1,1)$ mod 2, and we can then
deduce that necessarily $S=0$ for the latter states.

To prove the unicity, we still have to check whether only one
state exists at given $p$, $q$ and $s$. The argument which worked for the
$(p,q)$ states relied on the fact that any of these states can be
related to a singularity appearing on the Coulomb branch for some values
of the coupling, and that we know the multiplicity of the states
producing singularities. This clearly does not work for the
$(2p,2q)$ states, which are never massless (when $m\not =0$)
and thus never produce a singularity. As an alternative, we will rely
on a semiclassical analysis. In \cite{SZ} was shown that the states
$(2p,2q)$ indeed exist and are unique {\em for $|q|=1$.} The 
uniqueness of the $(2p,2q,S=0)$ states for {\em any} $p$ and $q$ 
relatively prime will then follow from the $\Gamma (2)$ invariance of 
the theory, see Section 8.
\section{Generalities about the curves of marginal stability}
Due to the BPS mass formula (\ref{BPSmass}), a BPS state is 
generically stable. Central charge as well as mass conservation indeed 
impose tights constraints. To see this, suppose that a BPS state of
central charge  $Z$ decay into $p$ states of central charges 
$Z_{1},\ldots ,Z_{p}$. This is possible only if
\begin{equation}
\label{firstc}
Z = Z_{1} + \cdots + Z_{p}
\end{equation}
and
\begin{equation}
\label{secondc}
|Z| = |Z_{1}| + \cdots + |Z_{p}|.
\end{equation}
A necessary condition for these equations to be compatible is that
\begin{equation}
\label{cms1}
\IM {Z_{k}\over Z}=0\enspace {\mathrm for}\enspace 1\leq k\leq p.
\end{equation}
Of the $p$ relations (\ref{cms1}), only $p-1$ are independent once
one takes into account (\ref{firstc}), and thus they define
a hypersurface of real codimension $p-1$ included in the Coulomb branch. 
In our case where the gauge group is of rank one and thus the Coulomb
branch of real dimension two, this means that decays into two distinct
particles can only occur along real curves, decays into three distinct
particles at special points, and decays into four or more distinct
particles are likely to be impossible.

Suppose we are studying the decay of a state $(\sigma ,s)$ into two 
states, one of them being $(\sigma ',s')$. Instead of using the 
original variables $\omega =(a_{D},a)$, it is convenient to use
shifted variables  $\Omega =(A_{D},A)$ such that $s=0$. The equation 
for the curve of marginal stability can then be cast in the following 
SL$(2,\Z )$ invariant form:
\begin{equation}
\label{cmseq}
{m\over\sqrt{2}}\, {\Omega\cdot\overline\Omega\over 2i
\IM\overline\Omega\cdot\sigma}={s'\over\sigma\cdot\sigma '}\cdotp
\end{equation}
The left hand side of (\ref{cmseq}) does not depend on the final 
state $(\sigma ',s')$, and the right hand side is an a priori 
arbitrary rational number which we will denote by $r$. We thus see 
that to a given particle $(\sigma ,s)$ is associated a whole family of 
curves of marginal stability, ${\cal C}_{(\sigma ,s)}(r)$,
indexed by a rational number $r$. 

It is useful at this stage 
to summarize what we need to show in order to complete 
our goal, which is to demonstrate that the spectrum of BPS states is in 
perfect agreement with $S$ duality both in the $N=4$ and massless 
$N_{f}=4$ theory, and also to prove the existence of the required 
states in the massive cases studied in Section 3.\\
--- We need to show that the states $(p,q)$, where $p$ and $q$ are 
relatively prime integers, which we have generated by analytic 
continuation in the {\em massive} theories (Section 4), still exist in 
the {\em massless} one. This will be established in Section 7 using the 
curves of marginal stability.\\
--- We need to show that the states $(2p,2q)$ exist in the $N_{f}=4$ 
theory, since they represent the duals of the W bosons in this case. 
We did not give any argument in favour of the existence of such states 
up to now, since they are not associated with any singularity on the 
Coulomb branch. However, one can establish their existence by using
the general argument presented in Section 8,
{\em and} by relying on the non trivial 
semiclassical results obtained in \cite{SZ} for states of magnetic 
charge 2.\\
--- Finally, we need to show that the dual of the  W bosons (and of 
the elementary quarks in the $N_{f}=4$ theory) does exist in the dual theory 
studied in Section 3.5. There it was only shown that the existence of 
these states does not lead to any inconsistency.
Actually, these dual states exist in the 
massless theories, and in Appendix B 
we will explain why they must still exist when the bare 
mass is increased and we follow the RG flow (\ref{RGdual})
toward the dual pure gauge (or massless $N_{f}=2$) theory.

Let us end this Section by mentioning the formula
\begin{equation}
\label{CPtrans}
\pmatrix{a_D\cr a\cr}\Bigl( \overline z\Bigr)=
\pmatrix{-1 & 2\cr 0 & 1\cr}
\pmatrix{\overline a_D\cr \overline a\cr}\Bigl(z\Bigr),
\end{equation}
which corresponds to a CP transformation. Note that though this is
{\em not} a duality transformation, it is still perfectly
compatible with the BPS mass formula (\ref{BPSmass}).
This relation is useful to study the symmetries and some
particular points of the curves of marginal stability.
\section{Existence of the states $(p,q)$}
In this Section, $p$ and $q$ are two relatively prime integers.
 
In Section 4, strong evidence in favour of the existence of the states
$(p,q)$ in the massless theories was given: depending on 
the value of the coupling $\tau $, when the bare mass is turned on, 
the singularities on the Coulomb branch may be due to any of these 
states. However, strictly speaking, the existence of the states was 
proven only when $m\not =0$ and $z=z_{j}$.
When $m=0$ and the singularities merge at $z=0$,
all the $(p,q)$ states become massless and thus degenerate in mass. 
Though a brutal discontinuity in the spectrum between $m\not =0$ and $m=0$ 
seems very unlikely on physical grounds, we would like to argue that 
the $(p,q)$ states not only exist at the singularities they are 
responsible for, but also at other points in the $z$ plane.
One way of doing this is to consider a configuration where the $(p,q)$ 
state becomes massless, and to study the family of curves of marginal 
stability associated with the decay of this state. If $(p,q)=(1,0)$ 
mod 2, we have a family ${\cal C}_{1}(r)$ defined by (cf (\ref{cmseq}))
\begin{equation}
\label{csmz1}
{\cal C}_{1}(r):\enspace\IM a_{D}(\overline a-{m\over\sqrt{2}})=
{m\over\sqrt{2}}\, r\, \IM a,\enspace r\in\Q
\end{equation}
and if $(p,q)=(1,1)$ mod 2 we have a family ${\cal C}_{3}(r)$,
\begin{equation}
\label{cmsz3}
{\cal C}_{3}(r):\enspace\IM a_{D}\overline a=
{m\over\sqrt{2}}\, r\, \IM (a-a_{D}),\enspace r\in\Q .
\end{equation}
The case of $(p,q)=(0,1)$ mod 2 is completely similar to the case 
$(p,q)=(1,0)$ mod 2 due to (\ref{periodstrans}).
We have computed the curves ${\cal C}_{j}(r)$
numerically, using the analytic formulas for the periods presented
in Appendix A, and the result is depicted in Figure 8.
\begin{figure}
\label{figure8}
\epsfxsize=14.7cm
\epsfbox{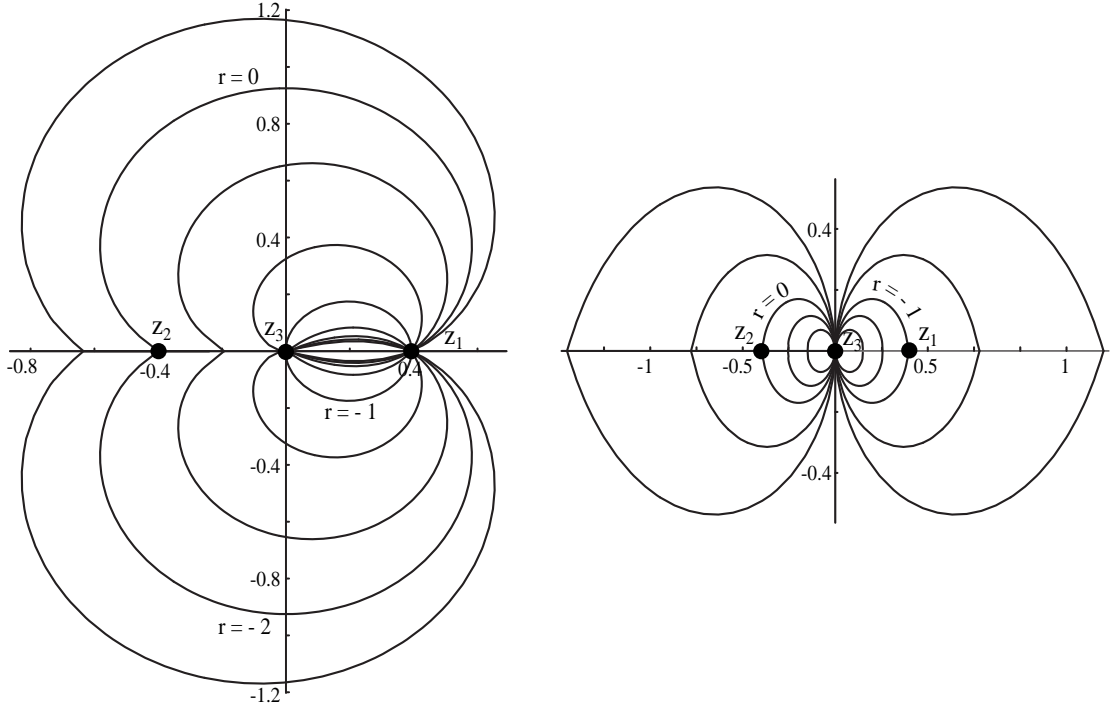}
\caption{The families ${\cal C}_{1}(r)$ (on the left)
and ${\cal C}_{3}(r)$ (on the right) for some 
values of $r$. The singularities at $z_{j}$ are represented by a bullet.
We chose $\tau =i$ and $m=1.5$.}
\end{figure}
It is easy to established, particularly in using (\ref{CPtrans}), that
the curves ${\cal C}_{3}(r)$ all intersect at $z_{3}$ and are
symmetric with respect to the real $z$ axis, while the curves
${\cal C}_{1}(r)$ all intersect at $z_{1}$ and $z_{3}$ and
$\overline {\cal C}_{1}(r)={\cal C}_{1}(-r-\sqrt{2}m)$.
More important, these curves do not intersect at any other points 
than the particular ones mentioned above and form a dense set in the 
whole $z$ plane. This is not prohibitive: the complementary set, 
generated by the curves ${\cal C}_{j}(x)$ for {\em irrational} values 
of $x$, is also a dense subset of the Coulomb branch, where the 
corresponding state is stable.
This shows that for any positive real number $R$,
there is always a point $z$ such that $|z|>R$ and which can be joined 
to the singularity $z_{j}$ by a curve ${\cal C}_{j}(x)$ where $x$ is 
taken to be irrational. As the state becoming massless at 
$z_{j}$ cannot decay along such a curve, it must exist at $z$.
As $R$ can be chosen independently of $m$, $(p,q)$ must exist in the 
$m=0$ theory for some $|z|>R$ and thus for any $z$.

To end this Section, I wish to point out that the preceding 
analysis suggests that states which cannot exist in the $m=0$ 
theory cannot exist anymore in the $m\not =0$ theory.
To understand this, assume that to the unwanted state 
$(p,q)_{s}$ is associated a point $z_{0}$ on the Coulomb branch such that
\begin{equation}
\label{unsing}
qa_{D}(z_{0})-pa(z_{0})+ms/\sqrt{2}=0,
\end{equation}
for any given $\tau $ and $m$.\footnote{Note that we did not 
prove this fact {\em at fixed} $\tau $,  though it seems to be true 
numerically.}.
The family of curves of marginal stability ${\cal C}_{(p,q)_{s}}(r)$
associated with $(p,q)_{s}$ looks like the families 
depicted in Figure 8, except that all the curves will intersect at 
$z_{0}$. Now suppose that $(p,q)_{s}$ exist at a point on a curve
${\cal C}_{(p,q)_{s}}(x)$ with $x$ irrational (such curves cover a 
dense subset ${\cal E}$ of the Coulomb branch). Then move 
$(p,q)_{s}$ along this 
curve, reach $z_{0}$ where it is massless and find an inconsistency
since $z_{0}$ is not a singular point. Thus $(p,q)_{s}$ cannot exist 
in the dense set ${\cal E}$: this means that it will never be a 
trully stable state in the theory.
Thus, if our hypothesis (\ref{unsing}) is correct, the
number of curves of marginal stability one must consider to study 
eventual decays in the massive theories is considerably decreased.
We will use this fact in Appendix B, see also [32]. 
\section{The general physical argument and $\Gamma (2)$ invariance}
In this Section, we wish to address a very general problem: when can 
we expect the monodromy group of a given theory to correspond to an 
exact duality symmetry? Of course, this a priori naive statement
is not always true, as 
Seiberg and Witten already pointed out in their original paper \cite{SWI} on 
the pure gauge theory. However, we would like it to be 
correct in the particular cases of the finite theories studied in this 
paper. We will give in the following some general physical arguments showing 
that this property
can be understood simply by looking at some very general 
features of the structure of the low energy effective action. Though 
we will restrict ourselves to very few examples, in the framework 
of SU(2) gauge theories, and chosen in order to 
illuminate the cases of SU(2) $N=4$ and $N=2$, $N_{f}=4$, the line of 
reasoning could a priori be applied to any $N=2$ theory.

First, let us consider a theory with only one singularity on the 
Coulomb branch. It could be the massless theories studied above.
Let us see why in this simple case,
the monodromy $M$ associated with this singularity must correspond 
to an exact duality symmetry of the theory. To the
singularity is associated a branch cut which extends to infinity and 
whose orientation clearly is arbitrary. By sweeping up the $z$ plane 
with such a branch cut, {\em while keeping all the parameters fixed,}
we show that necessarily two theories connected 
by the transformation $M$ must be physically equivalent. This kind of 
argument was already presented in subsection 4.3. 
In the case of $N=4$ or $N=2$ with four massless flavours, $M=-I$ and 
the ``duality'' transformation is nothing but a gauge transformation.

Let us now go to a more interesting situation, suited for instance for the
pure gauge theory, where two singularities are 
present at $z_{1}$ and $z_{2}$ with monodromies $M_{1}$ and $M_{2}$.
The theory now has an intrinsic scale $\Lambda $ such that 
$|z_{2}-z_{1}|=\Lambda ^{2}$. Two scenarios can be imagined in this case.
In the first one, for some reasons the naive arguments are true, and thus
the monodromy group generated by $M_{1}$ and $M_{2}$
corresponds to an exact duality symmetry. In the second scenario,
the two singularities combine: we choose the branch cut originating at
$z_1$ to go through $z_2$ and then to coincide with the branch cut 
originating at $z_2$. Far away from the singularities, we thus only see
a branch cut associated with a monodromy matrix $M_{\infty}$
which is the product of the two matrices $M_1$ and $M_2$. 
Of course the position of this branch cut extending to infinity can 
be chosen arbitrarily, and thus an exact duality symmetry must
be associated with $M_{\infty }$, at least when $|z|>>\Lambda ^2$.
In the case of the pure gauge theory, $M_{\infty}$ expresses the asymptotic
freedom and is associated with the invariance of the theory under
$\theta $ angle shifts.

Finally, let us study the case where three singularities $z_j$ are present
on the Coulomb branch, with monodromy matrices $M_j$. 
The theory may now have one, two, or three intrinsic scales, corresponding
to the distances $|z_j-z_k|$ between the singularities. In the
$N=2$ theory with one flavour of quark, we have three singularities
at equal distances from each other
because of a $\Z _3$ symmetry acting on the Coulomb branch
\cite{SWI,BF}. Thus we have only one scale $\Lambda $ at our
disposal. The same scenario as the one which prevailed for the
pure gauge theory occurs, and we do not have any particular duality
symmetry in this theory. Now, the finite theories
we are studying in this paper also have three singularities in the
Coulomb branch, but with the most important property that two
independent scales can be introduced and {\em any one of the
three singularities can be chosen to be arbitrarily far from the
other two.} Typically, we are studying a dual RG flow
as in (\ref{RGdual}), and two of the three
singularities (say $z_1$ and $z_2$) are separated by an almost fixed distance 
$\Lambda ^2$ while the third singularity ($z_3$)
runs to infinity with $m$ at a distance  $m^2$, where $m$ is 
the bare mass in the theory.
In such a configuration, as in the case where only two singularities
are present, we cannot know a priori whether the monodromies $M_1$ and $M_2$ 
associated with the singularities
at scale $\Lambda $ will correspond to an exact duality symmetry. 
However, the situation is different for the isolated singularity
at scale $m$. The latter can only be ``screened'' by the singularities
at scale $\Lambda $. We think that this is impossible, because it seems
very unlikely that the low energy physics at scale $\Lambda $ could
be influenced by a singularity at scale $m$, produced by a particle
which is ultra massive at scale $\Lambda $ (if it exists),
and vice-versa. One might argue that the branch cuts of the
three singularities could merge at a new intrinsic scale $\mu $. 
However, the physical origin of this new scale would be very
unclear and we will rule out this possibility.
We thus see that the only natural way to interpret
the singularity structure is to admit that $M_3$
corresponds to an exact duality symmetry of the theory at least in the
configuration where $|z_1-z_2|\sim\Lambda ^2$ and $z_3\sim m^2$ with
$m^2>>\Lambda ^2$. From the point of view of the massless theory, which
corresponds to $|z|>>m^2>>\Lambda ^2$, we have thus an invariance with 
respect to $M_3$. The r\^ole of the
three singularities can of course be exchanged due to the
SL$(2,\Z )$ invariance of the Seiberg-Witten curve
(\ref{curve}), and we deduce that
the $m=0$ theories have an exact duality group which contains
$\Gamma (2)$.

In particular, the existence of 
all the $(2p,2q)$ states in the $N_f=4$ theory
will then follow from the existence of the three 
states $(2,0)$ (the W), $(2,2)$ and $(0,2)$, which were
shown to exist in \cite{SZ} using the semiclassical method. 
\section{Conclusion}
In this paper, we proved rigorously that all the $(p,q)$ dyon states,
where $p$ and $q$ are relatively prime integers, exist in the
$N=4$ as well as in the $N=2$, $N_f=4$ gauge theories with gauge
group SU(2). These states were obtained by analytic continuation
of the $|q|\leq 1$ states which become massless at some points
of the Coulomb branch when the bare $\theta $ angle is zero.
In the process of analytic continuation, we saw explicitly that
the quantum numbers at the singularities can change according to
a monodromy transformation. 
We also proved the unicity of the dyon states, in perfect agreement
with an exact SL$(2,\Z )$ duality symmetry.
We also provided a very general physical argument showing
that $N=4$ and $N=2$ with four massless flavours must have an exact duality
symmetry group which contains the monodromy group of the corresponding
massive theories. This fact together with already known semiclassical
results implies the existence of all the $(2p,2q)$ vector states
in the $N_f=4$ theory.

The line of reasoning used in this work, particularly the
arguments of subsections 4.1 and 8, is likely to be 
generalizable to theories having any gauge group,
for instance to the case of SU(N) with $N_{f}=2N$. This may answer the
open questions in these cases \cite{controversy}.

From a very general point of view, our method may be seen as the physical 
counterpart of the string theory results based on the heterotic-type II 
duality conjecture \cite{VafLer}. 
It would be very interesting to prove the equivalence of the two
approaches. 
In our approach, one is seeking for non singular
points $z$ on the 
Coulomb branch and for cycles $c=(p,q)$ on the 
Seiberg-Witten curve $\Sigma _{z}$ such that
\begin{equation}
\label{conc1}
(c,\lambda )=\oint _{c}\lambda _{\mathrm SW} =0,
\end{equation}
where $\lambda _{\mathrm SW}$ is the Seiberg-Witten differential 
form, see Appendix A.
States $(p,q)$ corresponding to such a cycle $c$ cannot exist, and 
any other (hypermultiplet) states are very likely to exist.
In the string 
approach, stable states are associated with geodesics on $\Sigma 
_{z}$ endowed with the metric $|\lambda |^{2}$
where $\lambda $ is a particular element in the cohomology class of
$\lambda _{\mathrm SW}$. For states lying in a 
$N=2$ hypermultiplet, these geodesics must begin 
and end at the branching points $e_{j}$ (see (\ref{roots}) in Appendix A).
A link between the two points of view is the following: it is clearly 
impossible to have a non-null geodesic with the required properties for a 
state $c$ satisfying (\ref{conc1}), since the geodesic would have 
zero length and this requires its endpoints to coincide, that is
to have a singularity.
More important, the existence of a particular point $z$ such that 
(\ref{conc1}) is true is likely to prevent the existence of the 
corresponding geodesic for ``any'' $z$, by continuity. Of course 
subtleties are associated with the curves of marginal stability, and 
the vector particles may have a particular status. In any case, we 
believe that a complete understanding of the relationship between the 
two approaches is likely to provide a deeper understanding
of $N=2$ theories.
\ack
It is a pleasure to thank Adel Bilal for his permanent encouragements
and many interesting discussions. His critical reading of the
manuscript was also at the origin of many improvements.
I also thank Adam Schwimmer for a useful discussion, and
Aur\'elia who kindly processed the Figures.
\begin{appendix}
\vskip 1cm
{\bf Appendices}
\section{The computation of the periods}
\subsection{Notations and general formulas}
I present below some general formulas from the theory of elliptic
integrals (see, e.g., \cite{ellint}) in a form well suited for our 
purposes. By well suited, I mean first that they can be used 
straighforwardly to obtain the analytic solution for $a_{D}$ and $a$. 
Actually, the same general formulas can be used to evaluate the 
periods of the asymptotically free theories having $N_{f}\leq 3$ as 
well \cite{BFII}. Second, I mean that they can easily be implemented 
on a computer system,\footnote{The results will be expressed in
terms of special function already implemented in Mathematica.}
and thus can be used to determine numerically 
the curves of marginal stability.

Let us consider a torus whose equation is
\begin{equation}
\label{curvebis}
y^{2}=4\prod _{j=1}^{3}\bigl( x-e_{j}\bigr).
\end{equation}
In our particular case, it will be convenient to cast (\ref{curve}) into 
this form by performing the change of variable
\begin{equation}
\label{defxy}
x = X -E_{1}(\tau )z - {1\over 4}m^{2}E_{1}^{2}(\tau ),\enspace
y = 2Y.
\end{equation}
We will then have
\begin{eqnarray}
\label{roots}
e_{1}(z,\tau ,m)&=&0,\nonumber\\
e_{2}(z,\tau ,m)&=&\bigl( E_{2}(\tau )-E_{1}(\tau )\bigr) z+
{1\over 4}m^{2}\bigl( E_{2}^{2}(\tau )-E_{1}^{2}(\tau )\bigr), \nonumber\\
e_{3}(z,\tau ,m)&=&\bigl( E_{3}(\tau )-E_{1}(\tau )\bigr )z+
{1\over 4}m^{2}\bigl( E_{3}^{2}(\tau )-E_{1}^{2}(\tau )\bigr ).\\
\nonumber
\end{eqnarray}
The fundamental cycles on the torus (\ref{curvebis}) are chosen as 
indicated in Figure A.1.
\begin{figure}
\label{figure9}
\epsfxsize=14.7cm
\epsfbox{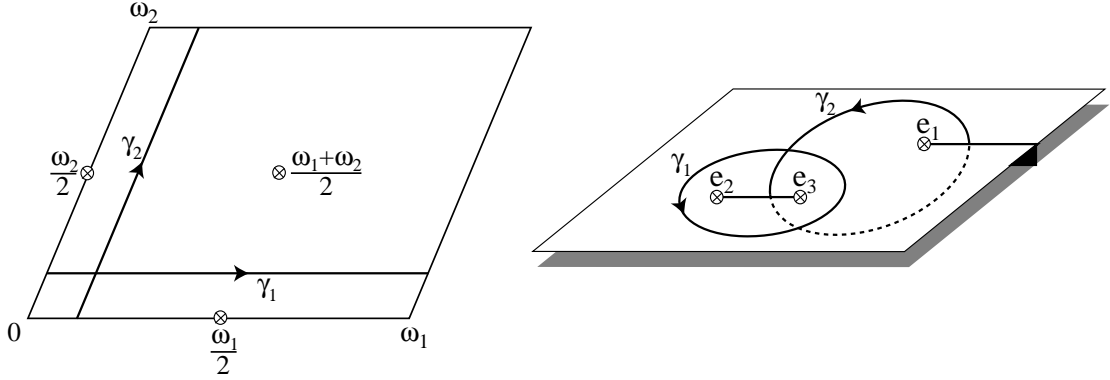}
\caption{The fundamental cycles $\gamma _{1}$ and $\gamma _{2}$ are 
represented in the two cases where the torus is viewed as a two 
sheeted surface with branch cuts from $e_{2}$ to $e_{3}$ and from
$e_{1}$ to $\infty $, or as a rectangle whose opposite edges are identified.}
\end{figure}
To this choice for the homology basis, we associate the usual 
variables $k$ and $k'$:
\begin{equation}
\label{defk}
k^{2} = {e_{3}-e_{2}\over e_{1}-e_{2}}\raise 
2pt\hbox{,}\quad
k'^{2}=1-k^{2}= {e_{1}-e_{3}\over e_{1}-e_{2}}\cdotp\\
\end{equation}
The fundamental periods are then given by
\begin{eqnarray}
\label{fundperiods}
\omega _{1}(z,\tau ,m) &=& \oint _{\gamma _{1}}{dx\over y}=
{2\over\sqrt{e_{1}-e_{2}}}\, K(k),\nonumber\\
\omega _{2}(z,\tau ,m) &=& \oint _{\gamma _{2}}{dx\over y}=
{2i\over\sqrt{e_{1}-e_{2}}}\, K(k'),\\
\nonumber
\end{eqnarray}
where $K$ is the standard complete elliptic integral of the first kind.
We will also need a formula involving elliptic integrals of the 
third kind. Maybe the 
most convenient form for this formula is the following:
\begin{equation}
\label{i3}
\oint _{\gamma _{j}}{dx\over (x-c)y}=
{1\over\wp '(u_{c})}\, \Bigl( 2\omega _{j}\, \zeta (u_{c})-4\eta _{j}\, 
u_{c} + 2i\ell\pi\Bigr),
\end{equation}
where $c$ is any complex number which is different from the roots $e_{j}$
and $\ell$ is an integer.
The $\eta _{j}$ are given by
\begin{eqnarray}
\label{eta}
\eta _{1}&=&-{1\over 2}\,\oint _{\gamma _{1}}{x\over y}\, dx =
\sqrt{e_{1}-e_{2}}\, E(k) - {e_{1}\over\sqrt{e_{1}-e_{2}}}\, 
K(k),\nonumber\\
\eta _{2}&=&-{1\over 2}\,\oint _{\gamma _{2}}{x\over y}\, dx =
-i\sqrt{e_{1}-e_{2}}\, E(k') - i{e_{2}\over\sqrt{e_{1}-e_{2}}}\, 
K(k')\\
\nonumber
\end{eqnarray}
in terms of the complete elliptic integrals of the first and second kinds 
$K$ and $E$.\footnote{Note that these functions may be expressed in 
terms of standard hypergeometric functions: 
$K(k)=(\pi /2) F(1/2,1/2,1;k^{2})$, $E(k)=(\pi /2) F(-1/2,1/2,1;k^{2})$.}
$\wp $ and $\zeta $ are straightforward 
generalizations (note that $e_{1}+e_{3}+e_{3}\not =0$ here)
of the Weirstrass $\wp $ and $\zeta $ functions,
and may be expressed in terms of the standard Jacobian and $\theta $
elliptic functions \cite{ellint} as
\begin{eqnarray}
\label{pzetadef}
\wp (u;z,\tau ,m) &=& e_{2}+{e_{1}-e_{2}\over\sn ^{2}
\bigl( (e_{1}-e_{2})^{1/2}u,k\bigr)}\nonumber\\
\wp ' (u;z,\tau ,m) &=& -2 (e_{1}-e_{2})^{3/2}\,
{\cn \bigl( (e_{1}-e_{2})^{1/2}u,k\bigr)\dn\bigl( (e_{1}-e_{2})^{1/2}u,k\bigr)
\over\sn ^{3} \bigl( (e_{1}-e_{2})^{1/2}u,k\bigr) }\nonumber\\
\zeta (u;z,\tau ,m) &=& {2\eta _{1}\over\omega _{1}}\, u +
{1\over\omega _{1}}\, {\theta ' \over\theta }\Bigl({u\over\omega 
_{1}}\Bigr).\\
\nonumber
\end{eqnarray}
Finally, $u_{c}$ is such that $\wp (u_{c})=c$, that is
\begin{equation}
\label{zcdef}
u_{c}=\wp ^{-1} (c)={1\over\sqrt{e_{1}-e_{2}}}\, \sn ^{-1}
\left[ \left( {e_{1}-e_{2}\over c-e_{2}} \right) ^{1/2},k\right].
\end{equation}
The indeterminate constant $\ell$ comes from the fact that the differential
form $dx/\bigl( (x-c) y\bigr) $ has poles with residues
$1/\wp '(u_c)$. If, under a continuous deformation, the integration
contour crosses such a pole, $\ell$ will be shifted by one unit.
Note also that the ambiguity that exists
in the definition of $u_{c}$, 
which stems from the fact that $\wp (u)$ is a doubly periodic, even 
function of $u$, can be absorbed in a redefinition of $\ell$ by
using Legendre's relation $\eta _1 \omega _2 - \eta _2 \omega _1 = i\pi $
and the fact that $\zeta(u+n_1\omega _1 + n_2 \omega _2)=
\zeta (u) +2(n_1 \eta _1+n_2 \eta _2)$ for any integers $n_1$ and $n_2$.
It should be clear that such integrals can only appear in theories
with non zero bare masses, since it is only here that
jumps in $a_{D}$ and $a$ are allowed \cite{SWII,FER}.
Changing $\ell$ will then simply amount to shifting the $s$ charge.

We are now equipped to compute the Seiberg-Witten periods $a_{D}$ and $a$ 
rather straighforwardly. However, two points remain to be 
discussed. First, when applying the above formulas, we are not 
certain to obtain a solution with the correct analytic structure
in the $z$ plane. 
This is explained in the next subsection where the effective coupling 
$t(z,\tau ,m)$ is computed. Second, we must find the Seiberg-Witten 
differential, if possible in a form where the canonical decomposition in 
terms of the building blocks treated above is simple. This is done in 
subsection 5.3. The solution is then given in 5.4.
\subsection{The effective coupling $t$}
The effective coupling $t$ is nothing but the modular parameter of the torus
(\ref{curve}), and would thus naively be given by $\omega _{2}/\omega _{1}$.
However, this formula is a priori valid only modulo SL$(2,\Z )$ 
transformations, and some additional constraints must be found from 
physics. The first constraint comes from the evaluation of $t$ in the 
limit $z\rightarrow\infty $ where it should coincide with $\tau $, modulo 
$\theta $ angle redefinitions which correspond to trivial shifts of the 
electric charge. This constraint is indeed satisfied with our choices for 
the homology cycles $\gamma _{1}$ and $\gamma _{2}$.
The second constraint comes from the analytic structure of $t$ as a 
function of $z$, which was described in Section 3 in the case of a trivial 
bare $\theta $ angle. 
This analytic structure is most easily studied when one uses the
explicit formulas
\begin{equation}
\label{explicitk}
k^2= {\theta _1 ^4 \over\theta _3 ^4}\,
{z-z_1\over z-z_3}\raise 2pt\hbox{,}\quad
k'^2= {\theta _2 ^4 \over\theta _3 ^4}\,
{z-z_2\over z-z_3},
\end{equation}
and
\begin{equation}
e_1-e_2 = (z-z_3)\,\theta _3 ^4,
\end{equation}
where the $z_j$ correspond to the positions of the singularities
(\ref{singpos}). We then obtain a unique solution, 
given when $\theta =0$ by
\begin{equation}
\label{solt}
t(z,\tau ,m)={da_D\over da}=
\left\lbrace
\begin{array}{lll}
\omega _2 /\omega _1 \quad &{\mathrm for}&\enspace
\IM z\geq 0.\\
\omega _2 /\omega _1 + 2\quad &{\mathrm for}&\enspace
\IM z < 0.\\
\end{array}
\right.
\end{equation}
$\partial a_D/\partial z$ and $\partial a/\partial z$ are also very
directly related to the fundamental periods $\omega _1$ and
$\omega _2$. Before writing explicit formulas, one must note
that though having the same monodromies, the solution for the
SO(3) and $N_f=4$ theories will differ by a global factor of
$1/2$ which directly comes from the analysis of the massless theories.
In the following, we will reserve the notation $(a_D,a)$ for the
solution of the SO(3) theory, and for $N_f=4$ 
we will have $(a_D^{N_f=4},a^{N_f=4})=(a_D/2,a/2)$. With these
conventions,
\begin{equation}
\label{solda}
{\partial a\over \partial z}=
{\sqrt{2}\over 2\pi}\,\omega _1 = 
{\sqrt{2}\over \pi} {1\over\theta_{3}^{2}\sqrt{z-z_{3}}}\,
K(k)
\end{equation}
and
\begin{equation}
\label{soldad}
{\partial a_{D}\over\partial z}=
\left\lbrace
\begin{array}{lll}
{\displaystyle\sqrt{2}\over\displaystyle\pi}\,
{\displaystyle 1\over\displaystyle\theta_{3}^{2}\sqrt{z-z_{3}}}\,iK(k')
\quad &{\mathrm for}&\enspace\IM z\geq 0.\\
{\displaystyle\sqrt{2}\over\displaystyle\pi}\,
{\displaystyle 1\over\displaystyle\theta_{3}^{2}\sqrt{z-z_{3}}}\,
\bigl( iK(k') + 2K(k) \bigr)
\quad &{\mathrm for}&\enspace\IM z < 0.\\
\end{array}
\right.
\end{equation}
\subsection{The Seiberg-Witten differential}
We now wish to integrate the relations (\ref{solda},\ref{soldad}) in
order to obtain $a_D$ and $a$. To do this, the standard method is
to find a one form $\lambda $ (the so-called Seiberg-Witten differential)
so that
\begin{equation}
\label{eqdifl}
{\partial\lambda\over \partial z}={dx\over dy}
\end{equation}
up to an exact form. We will then compute $a_D$ and $a$ by mean
of the formulas
\begin{equation}
\label{periodint}
a (z,\tau ,m)={\sqrt{2}\over 2 \pi }\oint _{\gamma _1} \lambda ,
\quad
a_D (z,\tau ,m)={\sqrt{2}\over 2 \pi }\oint _{\gamma _2} \lambda .
\end{equation}
The simplest way to find $\lambda $ certainly is to use,
instead of the variables $(y,x)$, the rescaled
variables $(\tilde y=y/(2\theta _{2}^{3}\theta _{3}^{3}),\,
\tilde x= x/ \theta _{2}^{2}\theta _{3}^{2})$\footnote{This 
rescaling of the
variables is of course innocent from the point of view 
of (\ref{eqdifl}). We perform it in order to simplify the expressions.
On the contrary, the use of $(\tilde y,\tilde x)$ instead of
$(Y,X)$ amounts to adding a non trivial exact differential to $\lambda $,
which simplifies its form nicely.}
in terms of which the curve takes the form
\begin{equation}
\label{curvebis2}
\tilde y^{2}=\tilde x M(\tilde x,z,\tau ,m)N(\tilde x,z,\tau ,m)
\end{equation}
with
\begin{equation}
{\partial\tilde y^{n}\over\partial z}={1\over 2}n\tilde y^{n-2}\tilde x (M+N).
\end{equation}
By multiple integration by part we then obtain
\begin{eqnarray}
\int {dz\over \tilde y}&=&\int dz\, {\tilde xR\over \tilde y}\, {1\over \tilde xR}
={\tilde y\over \tilde xR} +
\int dz\, (3\tilde y\tilde xR) {1\over 3\tilde x^{2} R^{3}}=\cdots\nonumber\\
&=& {1\over\sqrt{x}}\sum _{n=1}^{\infty}{1\over 2n-1}\,
\left( {y\over\sqrt{x} R} \right)^{2n-1}=
{1\over 2\sqrt{x}}\ln {\sqrt{x}R+y\over\sqrt{x}R-y}\cdotp\\
\nonumber
\end{eqnarray}
A very convenient and simple form for $\lambda $ will then be
\begin{equation}
\label{swdiff}
\lambda = -{\sqrt{x}\over\theta _{2}\theta _{3}}
\, d\ln {\sqrt{x}R+y\over\sqrt{x}R-y}
= \Bigl( z- z_{1}\Bigr)\, {2x\over y}{dx\over x+{1\over 4}m^{2}\theta 
_{2}^{4}\theta _{3}^{4}}.
\end{equation}
\subsection{The computation of $a$ and $a_D$}
We end our computation by writing $\lambda $ in the form
\begin{equation}
\lambda = 2\bigl( z-z_{1}\bigr) {dx\over y} - {1\over 2}m^{2}
\theta _{2}^{4}\theta _{3}^{4} \bigl( z-z_{1}\bigr) 
{dx\over y\bigl( x + {1\over 4}m^{2}\theta _{2}^{4}\theta 
_{3}^{4}\bigr)}\cdotp
\end{equation}
By noting that
\begin{equation}
\wp '^{2}(u)=4\prod _{j=1}^{3}\Bigl( \wp (u) -e_{j} \Bigr),
\end{equation}
we can easily compute $\wp '(u_{c})$ with $c=-m^{2}\theta 
_{2}^{4}\theta _{3}^{4}$ and then apply (\ref{fundperiods}, \ref{i3}) 
to find the desired solution:
\begin{equation}
\label{sola}
a(z,\tau ,m)={\sqrt{2}\over\pi }\Bigl[ \bigl( z-z_{1} \bigr)\, \omega _{1}
+ {i\over 4}m\, \bigl( 2\omega _{1}\zeta (u_{c}) -4\eta _{1}u_{c} 
\bigr) \Bigr],
\end{equation} 
and
\begin{eqnarray}
\label{solad}
a_{D}(z,\tau ,m)&=&\left\lbrace
\begin{array}{l}
{\displaystyle\sqrt{2}\over\displaystyle\pi }
\Bigl[ \bigl( z-z_{1} \bigr)\, \omega _{2}
+ {i\over 4}m\, \bigl( 2\omega _{2}\zeta (u_{c}) -4\eta _{2}u_{c} 
\bigr) \Bigr]  \quad {\mathrm for}\enspace
\IM z\geq 0.\\
{\displaystyle\sqrt{2}\over\displaystyle\pi }
\Bigl[ \bigl( z-z_{1} \bigr)\, (\omega _{2}+2\omega _{1})
+ {i\over 4}m\, \bigl( 2(\omega _{2}+2\omega _{1})\zeta (u_{c})
-4(\eta _{2}+2\eta _{1})u_{c} 
\bigr) \Bigr]\\
\end{array}\right.\nonumber\\
& &\hskip 8cm {\mathrm for}\enspace
\IM z < 0.\\
\nonumber
\end{eqnarray}
The different prescriptions for $\IM z \geq 0$ or for $\IM z < 0$ 
correspond of course to a zero $\theta $ angle. The generalization to 
any $\theta $ angle should be clear in view of the discussion in 
Section 4.
\section{Duality in the massive theories}
\begin{figure}
\label{figure10}
\epsfxsize=14.7cm
\epsfbox{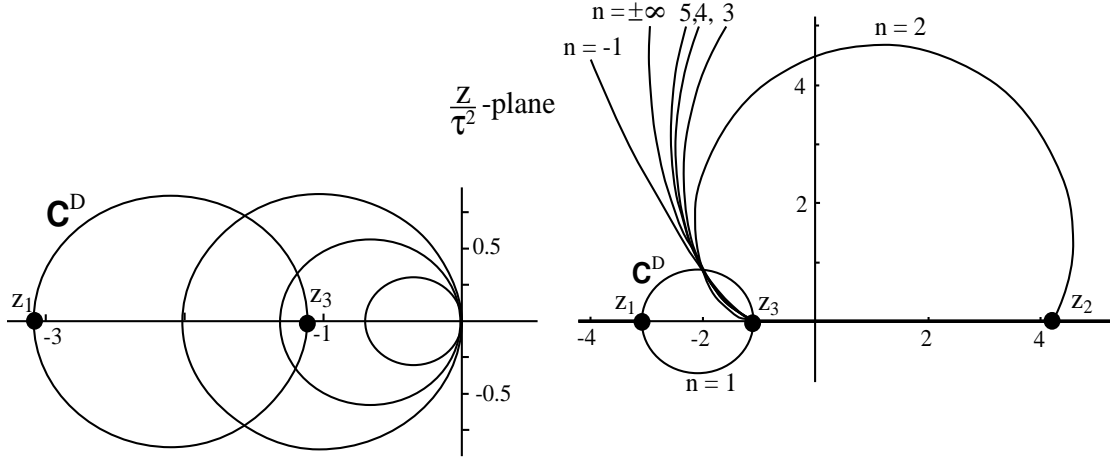}
\caption{The curves of marginal stability corresponding to the decay of the
state $(1,0)_{-1}$ (or $(2,0)_{-2}$). On the left are represented the curves
${\cal C}^D$ (n=1) for $\tau =i$ and various values of the mass $m$
from 0 to the RG value given by (\ref{RGdual}), and then for 
$m=5$ and $\tau $ given by (\ref{RGdual}). On the
right, curves for $m=5$ and $\tau $ given by (\ref{RGdual}) are represented
for various values of $n$.}
\end{figure}
In this Appendix, we complete the proof of Section 3 and show that
the spectrum of the dual pure gauge theory (or dual $N_f=2$ massless theory)
is in perfect agreement with duality. 
In Section 3, we already showed that the duals of the $|n_m|=1$ monopoles
indeed exist outside the curve of marginal stability ${\cal C}^D$ represented
in Figure 4 in the theory $(\tau ^2 z,-1/\tau ,m)$.
In addition to these states, duality predicts the existence
of the duals of the W bosons, and also of the elementary quarks in the case of
the $N_f=4$ theory. More precisely, we need to prove that
in the SO(3) theory
the state $(0,1)_{-1}$ indeed exists for $\IM z<0$ outside the curve 
${\cal C}^D$ or equivalently that $(2,1)_1$ exists for $\IM z>0$.
In the $N_f=4$ theory, in addition to the preceding state (now interpreted
as being the dual of the quarks of the massless $N_f=2$ theory), 
we must have
the $(0,2)_{-2}$ (or $(4,2)_2$) state interpreted as being the dual of the W.

Our strategy will be the following: as we already know that all these
states exist when $m=0$, we will simply show that they cannot decay 
outside the curve ${\cal C}^D$ when
the mass is turned on and we follow the RG flow (\ref{RGdual}) 
toward the dual pure
gauge (or $N_f=2$) theory. To do this, remark first that the family of curves
of marginal stability associated with $(0,1)_{-1}$ coincides with the
one for $(0,2)_{-2}$, and that the family for $(2,1)_{1}$ is the
symmetric of the family for $(0,1)_{-1}$ with respect to the real $z$ axis.
Thus, we need to study only one family of curves, whose defining equations
deduced from (\ref{cmseq}) can be cast in the form
\begin{equation}
\label{curveap}
\IM \Bigl(a_D-{m\over\sqrt{2}}\Bigr)\Bigl(\overline a-{m\over\sqrt{2}}\Bigr)=
\Bigl( {S'\over n_e'}-1\Bigr) \, {m\over\sqrt{2}}\, \IM a_D.
\end{equation}
In this equation, $S'$ represents the physical $S$ charge a state
produced in the decay reaction would have when $m=0$, and $n_e'$
represents the electric charge of the same state. As argued in
Section 7, we will consider only $S'$ such that 
$|S'|\leq 1$ and thus $S'/n_e'=0$
or $S'/n_e'=1/n$ where $n$ is any integer. The corresponding curves 
are represented in Figure B.1. As there is no curve outside
${\cal C}^D$ in the lower half-plane, the state $(0,1)_{-1}$ (or
$(0,2)_{-2}$) cannot decay in this region. This completes the proof.
\end{appendix}
\end{document}